\documentclass[useAMS,usenatbib]{mn2e}
\usepackage{epsfig,amssymb}
\usepackage{graphicx,amsmath,multicol}
\usepackage{multirow}
\usepackage{caption}
\usepackage{subcaption}
\usepackage{mathrsfs}

\title[Super-Chandrasekhar white dwarfs in GRMHD]{GRMHD formulation of highly super-Chandrasekhar rotating magnetised white dwarfs: Stable configurations of non-spherical white dwarfs}
\author[S. Subramanian \& B. Mukhopadhyay]
{Sathyawageeswar Subramanian\thanks{sathynius2@gmail.com} and Banibrata Mukhopadhyay\thanks{bm@physics.iisc.ernet.in} \\ Department of Physics, 
Indian Institute of Science, Bangalore, India 560012}
\begin{document}

%%% sathya@ug.iisc.in

\date{Accepted Received ; in original form }

\pagerange{\pageref{firstpage}--\pageref{lastpage}} \pubyear{2015}

\maketitle

\label{firstpage}

%%%%%%%%%%%%%%%%%%%%%%%%%%%%%%%%%%%%%%%%%%%%%%%%%%%%%%%%%%%%%%%%%%%%%%%%%%%%%%%

\begin{abstract}
Here we extend the exploration of significantly super-Chandrasekhar magnetised white dwarfs by numerically computing axisymmetric stationary equilibria of differentially rotating magnetised polytropic compact stars in general relativity (GR), within the ideal magnetohydrodynamic regime. We use a general relativistic magnetohydrodynamic (GRMHD) framework that describes rotating and magnetised axisymmetric white dwarfs, choosing appropriate rotation laws and magnetic field profiles (toroidal and poloidal). The numerical procedure for finding solutions in this framework uses the $3+1$ formalism of numerical relativity, implemented in the open source {\it XNS} code. We construct equilibrium sequences by varying different physical quantities in turn, and highlight the plausible existence of
super-Chandrasekhar white dwarfs, with masses in the range of $2-3$ solar mass, with central (deep interior) magnetic fields of the order of $10^{14}$G and differential rotation with surface time periods of about $1-10$ seconds. We note that such white dwarfs are candidates for the progenitors of peculiar, overluminous type Ia supernovae, to which observational evidence ascribes mass in the range $2.1-2.8$ solar mass. We also present some interesting results related to the structure of such white dwarfs, especially the existence of polar hollows in special cases.

\end{abstract}

%%%%%%%%%%%%%%%%%%%%%%%%%%%%%%%%%%%%%%%%%%%%%%%%%%%%%%%%%%%%%%%%%%%%%%%%%%%%%%%


\section{Introduction}

Recently, there has been a lot of interest in exploring the possible existence of
massive compact objects, in particular neutron stars and white dwarfs, based on the theory
of stellar structure. This is primarily because of the observational evidences for 
massive neutron star binary pulsars PSR~J1614-2230 (\citealt{nat}) and  PSR~J0348+0432 (\citealt{sc}) 
with masses 
$1.97M_\odot$ and $2.01M_\odot$ respectively, where $M_\odot$ is the mass of the Sun. 
\cite{weissenborn} examined the possibility of very massive neutron stars in the presence of 
hyperons and the conditions to obtain the same. \cite{whittenbuary}, based on 
quark-meson coupling model, showed that the maximum mass of neutron stars could be $\approx 2M_\odot$,
when nuclear matter is in $\beta$-equilibrium and hyperons must appear. Apart from the exploration
based on the equation of state (EoS), as mentioned above, neutron stars with mass $\gtrsim 2M_\odot$ were shown to be possible
by exploring effects of magnetic fields, with central field $B_c\sim 10^{16}$G (\citealt{pili,monica}), and modification to Einstein's gravity (\citealt{eksi,capo,cheon}).

On the other front, the discovery of over-luminous type~Ia supernovae 
(e.g. SN~2003fg, SN~2006gz, SN~2007if, SN~2009dc) argues for significantly super-Chandrasekhar 
progenitors, of mass $2.1-2.8M_\odot$, in such supernovae (e.g. \citealt{nature,hicken,yam,scalzo,silverman,taub}), 
which further argues for the super-Chandrasekhar limiting mass of white dwarfs
as a potential possibility. Mukhopadhyay and his collaborators initiated the exploration of the possible 
existence of such a super-Chandrasekhar limiting mass, first by exploiting the effects of high 
magnetic fields in white dwarfs (e.g. \citealt{kundu,prd12,ijmpd12,prl13,grf13,apjl13,mpla14,jcap14,jcap15a}), 
and subsequently exploring the effects of modifying Einstein's gravity (\citealt{jcap15b}).

In fact, the existence of super-Chandrasekhar white dwarfs was shown to be possible many decades back. 
Ostriker and his collaborators attempted to model magnetised and/or rotating 
white dwarfs in the Newtonian framework in a series of papers (\citealt{Ost67,Ost68a,Ost68b,Ost68c,Ost69,Ost70,Ost71}). The maximum mass of a non-magnetised Newtonian white dwarf was reported then to be $2.26M_\odot$,
with, however, equatorial radius $2510$km (\citealt{Ost69}).
A couple of decades later, \citet{adam} obtained significantly super-Chandrasekhar Newtonian magnetised
white dwarfs, but did not impose realistic constraints on the magnetic fields.
Nevertheless, none of them could pinpoint whether there is a limiting mass or not, in the same spirit as 
the Chandrasekhar mass-limit
for non-magnetised white dwarfs was proposed (\citealt{chandra35}).
However, \citet{prl13} showed the existence of a new mass-limit
for white dwarfs in the presence of magnetic field, but in the absence of rotation, with a simpler (constant) 
field configuration.
In fact, after the initiation by the Mukhopadhyay-group, recently the topic of 
super-Chandrasekhar white dwarfs has been given renewed attention by the community.

%Although some concerns were {\it speculatively} raised against the proposition of the Mukhopadhyay-group
%(\citealt{chamel,nityakon,ruffini}), 
%they were answered and addressed in detail in subsequent works (\citealt{mpla14,jcap14};  \citealt{vishal}), and were shown 
%(\citealt{reply,chin}) some of them (e.g. \citealt{nityakon}) just to be erroneous. However,
%some other authors independently supported the proposal of magnetised super-Chandrasekhar white dwarfs
%based on their respective computations (\citealt{cheon,herrera1,herrera2}).

Nevertheless, apart from magnetic fields, rotational effects also could increase the mass of white dwarfs above the Chandrasekhar-limit,
as was shown by \citet{Ost68c,ruff1}. In fact, in general, a white dwarf could be rotating as well as magnetised. Its angular frequency, however, conventionally, is expected to be much smaller than that of a typical neutron star.
Nevertheless, in the presence of high rotation, white dwarfs are expected to be oblate spheroids. On the other hand, 
in the presence of poloidally and toroidally dominated magnetic fields, the respective white dwarfs
are expected to be oblate and prolate spheroids. Therefore, the overall shape of rotating magnetised white dwarfs
depends on the relative effects of rotation and the components of magnetic field.

Recently, \citet{jcap15a} have explored, using the (modified) {\it XNS} 
code, general relativistic magnetohydrodynamic (GRMHD) 
analyses of magnetised but nonrotating white dwarfs and showed that their mass 
could be as high as $3.4M_\odot$.
In this paper, we explore GRMHD numerical modeling of 
white dwarfs, including both magnetic and rotational effects, 
by using the {\it XNS} code. Note that the {\it XNS} code
is particularly suited to solving for highly deformed stars due to a strong 
magnetic field and/or rotational effects (\citealt{xns1,pili}). 
Although originally the {\it XNS} code was developed to investigate neutron stars, recently it was
modified \citep{jcap15a} in order to obtain
equilibrium configurations of deformed, magnetised white dwarfs.
Here, we further modify it to make it suitable for deformed, magnetised, differentially rotating white dwarfs.

This paper is organized as follows. In Section 2, we briefly mention the
numerical approach as described by \citet{jcap15a}. In section 3, we describe the organisation of the results
to be presented in the subsequent sections and chosen basic features/assumptions and units for the work. 
In section 4, we discuss the results of non-magnetised, differentially rotating white dwarfs. In sections 5 and 6, we present the results for differentially rotating white dwarfs having toroidal and poloidal magnetic field 
configurations respectively. We also explore the possibility of magnetised, rotating white dwarfs 
having very small radii, in section 7. We finally conclude in section 8 with a summary.

%%%%%%%%%%%%%%%%%%%%%%%%%%%%%%%%%%%%%%%%%%%%%%%%%%%%%%%%%%%%%%%%%%%%%%%%%%%%%%%

\section{Numerical approach and set-up}

We refer the readers to \citet{xns1}, \citet{pili} and \citet{jcap15a}
for a complete description of the GRMHD equations and numerical approach.
Our interest is obtaining equilibrium solutions of high density, rotating magnetised, 
relativistic white dwarfs. 

For the convenience, we recall below the equation pertaining to the differential rotation profile chosen in this paper. Stationary configurations can have many rotation laws consistent with the hydrostatic balance condition in GRMHD. One such form, deduced based on the Newtonian limit (\citealt{Ster03}), is used in {\it XNS}, given by

% =\frac{\left(\Omega-\omega\right)\Phi^2r^2\text{\,sin}^2\theta}{N^2-\left(\Omega-\omega\right)\Phi^2r^2\text{\,sin}^2\theta},

\begin{equation}
j(\Omega)=A^2\left(\Omega_c-\Omega\right),
\label{jProfile}
\end{equation} where $j$ is related to the specific angular momentum, $\Omega$ is the angular velocity of a local zero angular momentum observer (ZAMO) as measured from infinity, $\Omega_c$ is the angular velocity at the centre of the coordinate system, and $A$ is a positive constant that is implicitly related to $\Omega_c$. We will refer to $A$ as the differential rotation parameter. When 
$\Omega \to \Omega_c$, $A \to \infty$ and the system is of uniform rotation. In the opposite limit, when $A \to 0$, $j$ becomes constant in space. Finite values of $A$ indicate a spatial variation of angular velocity, and hence differential rotation. 

%
%The magnetic field profile that ensures integrability of the hydrostatic balance equation of GRMHD is a type of magnetic barotropic law, in which $\alpha RB$ is a function of $\alpha^2R^2h\rho$ only. The XNS code uses a polytropic law of the form
%\begin{equation}
%\alpha RB = K_m(\alpha^2R^2h\rho)^m,
%\label{magProfile}
%\end{equation} where $m \geq 1$ (otherwise the magnetic energy integral diverges) is called the magnetic polytropic index.
%

%%%%%%%%%%%%%%%%%%%%%%%%%%%%%%%%%%%%%%%%%%%%%%%%%%%%%%%%%%%%%%%%%%%%%%%%%%%%%%%%

\section{Organization of results}
We have grouped the various sequences broadly under three sections --- non-magnetised rotating sequences, sequences with toroidal magnetic field, and sequences with poloidal magnetic field.

\begin{figure}
\centering
\begin{subfigure}[h]{0.45\linewidth}
\includegraphics[width=\linewidth]{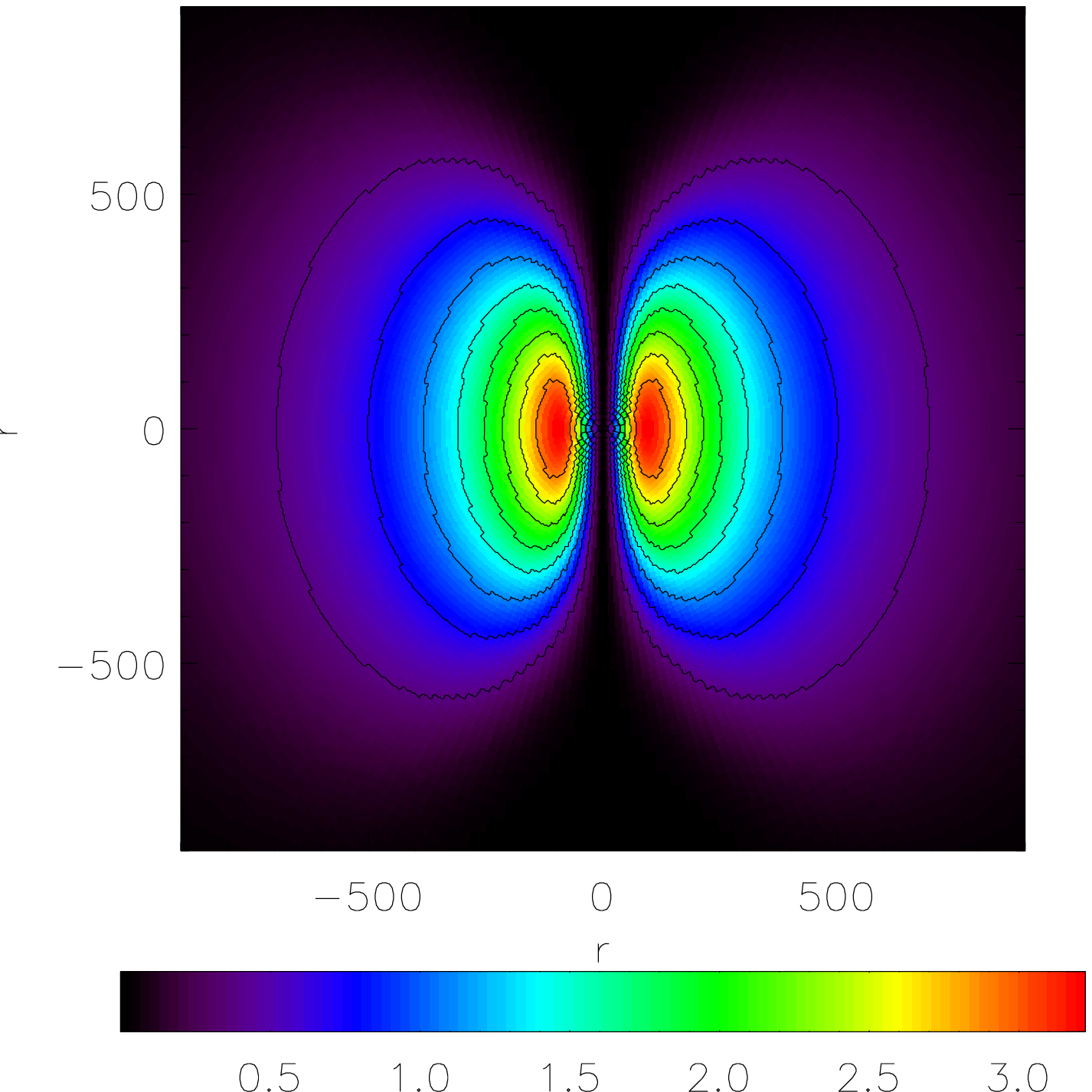}
\caption{\large\bfseries(a)}
\end{subfigure}
\begin{subfigure}[h]{0.45\linewidth}
\includegraphics[width=\linewidth]{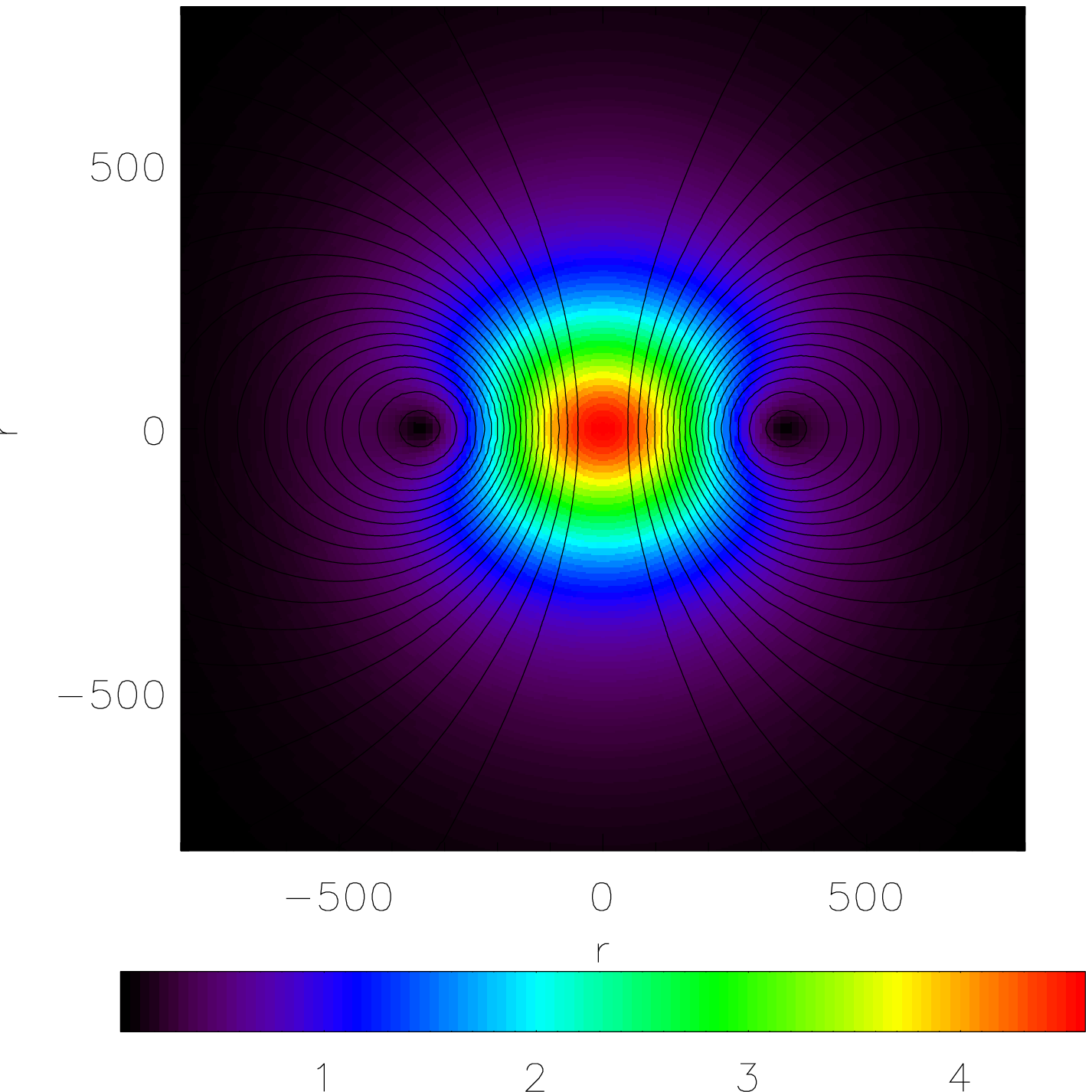}
\caption{\large\bfseries(b)}
\end{subfigure}
\caption[Magnetic field geometries]{Illustration of magnetic field ($B$) geometries: The panels are contour plots of the magnetic surfaces, and $B$ in units of $10^{14}$G for (a) purely toroidal field, and (b) purely poloidal field, 
when $r$ is in the units of 1.48km.}
\label{magGeom}
\end{figure}

At the outset, we illustrate the two kinds of field geometries in Fig. 
\ref{magGeom}, which will be considered throughout. It may be noticed that the toroidal field is zero at the centre and peaks 
at a small radius, distance $r$ from the centre, within the star. All the chosen toroidal field profiles, 
throughout the paper,
correspond to $m=1.4$, unless stated otherwise, 
where the number $m$ is the toroidal magnetisation index that determines the form of the barotropic magnetic 
field profile (see, e.g., \citealt{xns1}).
The poloidal field we choose is purely dipolar, having no non-linear current sources. Various 
field profiles, chosen in the paper, are determined by our respective choice of the maximum value of magnetic field
and the corresponding magnetic dipole moment obtained for the self-consistency of the solutions (see \citealt{pili,
jcap15a}, for details).

We compute all our sequences at the central density $\rho_c = 1.9902\times 10^{10}$g/cc, which is high enough to ensure almost complete relativistic electron degeneracy, so that we may use the polytropic EoS with $n=3$ (i.e. $\Gamma=\frac{4}{3}$) consistently throughout the star. This density is still lower than the limit at which gravitational instabilities in GR set in, which is about $3\times 10^{10}$g/cc.

\begin{figure}
\centering
\includegraphics[width=0.6\linewidth]{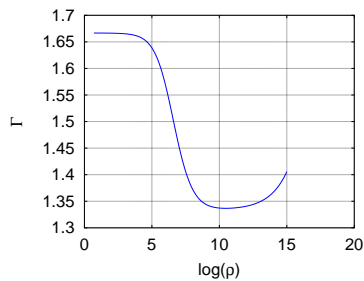}
\caption[Polytropic $\Gamma$ that locally fits the Chandrasekhar EoS, as a function of density]{Polytropic 
$\Gamma$, that locally fits the Chandrasekhar EoS for a zero temperature electron gas, as a function of density in 
g/cc.}
\label{chandraG}
\end{figure}

The choice of a polytropic EoS with $n=3$ throughout is almost appropriate for white dwarfs, 
as can be seen from Fig. \ref{chandraG}, which shows the best polytropic 
$\Gamma$, that fits approximately the full zero temperature EoS locally, at various densities. It is clear from the figure that $\Gamma=4/3$ gives a good fit in a 
wide density range, from about $10^{7}$ to $10^{13}$g/cc. Since the outer, low density, layers of a white dwarf do not contribute much to the global properties such as total mass, we can safely use this model to make conclusions about such properties of white dwarfs. 

Hence, all the sequences described in this paper have been computed with the 
fixed EoS 
\begin{equation}
P=k_3\rho^{\frac{4}{3}},
\label{EoS} 
\end{equation} where $k_3=4.9018\times 10^{14} cc\; s^{-2}\; g^{-1/3}$. 
%at the fixed central density $\rho_c=1.9902\times 10^{10}$g/cc. 

In the discussion and all the tables and figures that follow, the units used are the following, unless mentioned 
explicitly.
\begin{enumerate}
\item Mass - solar mass ($M_\odot$); $M_\odot = 1.987\times 10^{33}$g.
\item Radius - km.
\item The radial coordinate in all the contour plots is in units of 
$r_0=1.48$km (chosen because the normalisation $G=c=M_\odot=1$ has been followed).
\item Magnetic fields ($B$) - $10^{14}$G.
\item Angular velocities ($\Omega$) - rad/s.
\item The normalisation factor of density for the contour plots of $\log(\rho/\rho_0)$ is $\rho_0=10^{10}$g/cc.
\end{enumerate}
All the chosen differential rotations (\citealt{KEH1,xns1}) correspond to $A^2=10^5$, unless stated otherwise. 

Numerical solution of equations involves discretisation of the domain. We have used a resolution of $N_r=250$ for the radial coordinate, and $N_\theta=100$ for the angular coordinate. The {\it XNS} solver uses a compactified domain, so that the range for the radial coordinate always starts at 0 and truncates at some specified $R_{max}$. For the present sequences, $R_{max}$ varies from 1500km to 4000km. It may be noted that for numerical purposes, the pressure and density at the surface of the star are not zero, but a small, finite value at which the integration is truncated. 

The choice of grid, we have followed, is drawn from the guide to the {\it XNS} code.\footnote{http://www.arcetri.astro.it/science/ahead/XNS/files/Guide.pdf} The angular coordinate varies from $\theta=0$ to $\pi$. 
The choice of 100 angular grid points appears to be sufficient to capture the variation of quantities with 
$\theta$ for the models presented here, which have no cusps or other sharp features. 
The value of $R_{max}$ can, in principle, be arbitrarily chosen. However, it is necessary 
to guarantee that the white dwarf is well resolved over a sufficient number of grid points (50-100), 
so this parameter and $N_r$ need be chosen consistently. Our choice of $N_r=250$ ensures that all the 
configurations we compute have well over 150 grid points within the stellar surface. 
The Tolman-Oppenheimer-Volkoff (TOV) equation solver routine of {\it XNS}, which computes the initial guess 
TOV solution for the metric solver, converges when the ADM masses measured at $R_{max}$ and $R_{max}/2$ 
coincide within a given tolerance. Furthermore, the iterative metric solver of {\it XNS}, based on conservative 
variables, measures convergence by the quantity ${\hat D} = \psi^6\rho_c$, which contains the unknown value 
of the conformal factor $\psi$ at the centre. All the results presented in this paper correspond to convergent 
models and are stable at the selected resolution.

We also recall at the outset that the non-magnetised, non-rotating spherical equilibrium configuration of a white dwarf with the central density 
$\rho_c=1.9902\times 10^{10}$g/cc has mass $M=1.416$ and
radius $r_e=1215$ (\citealt{jcap15a}).

%\clearpage 	% clear the floats

%%%%%%%%%%%%%%%%%%%%%%%%%%%%%%%%%%%%%%%%%%%%%%%%%%%%%%%%%%%%%%%%%%%%%%%%%%%%%%%

\section{Non-magnetised rotating configurations}
We start with the sequences of non-magnetised rotating stars, focussing on 
differentially rotating configurations. Uniform rotation has a marginal effect 
on the mass and radius of the star, except close to the mass-shedding limit. As pointed out by \citet{Ost68a}, uniformly 
rotating sequences are terminated at low values of the kinetic to
gravitational energy ratio (KE/GE=0.05-0.06), by the balancing of gravitational and centrifugal accelerations at the surface. However, uniform rotation does result in severe distortion of the outer layers.

\cite{Ost70} arrived at a criteria for stability of rotating stars based on the value of KE/GE, by showing that configurations become unstable to non-axisymmetric perturbations when KE/GE $\geq 0.27$. \citet{KEH1} have adopted a similar criterion for stability of rotating configurations, and argued that when relativistic effects are stronger and ring-like or toroidal structures form, the KE/GE threshold reduces to $\approx 0.14-0.16$. 
With this in mind, we have always terminated our sequences at KE/GE $\leq 0.14$,
 even when the code is capable of converging to an equilibrium solution for 
higher values of this ratio. 

We plan to explore the phenomenon of `polar hollows'. The term polar hollow refers to a change 
of the density isocontours of the star from a convex to a concave nature, near its poles along the rotation axis.  Polar hollows in polytropic stars have previously been reported by \citet{KEH2}, and investigated in detail for the Newtonian case by \citet{polHol}. To the best of our knowledge, not much work has been done on polar hollows in rotating and magnetised stars in GR. In this work, we take a first step towards understanding this problem, by computing and demonstrating the plausible existence of equilibrium solutions with polar concavities, for the specific angular momentum distribution profile given by 
equation (\ref{jProfile}).

\subsection{Differentially rotating sequence with varying $\Omega_c$}
\begin{figure}
\centering
\includegraphics[width=\linewidth]{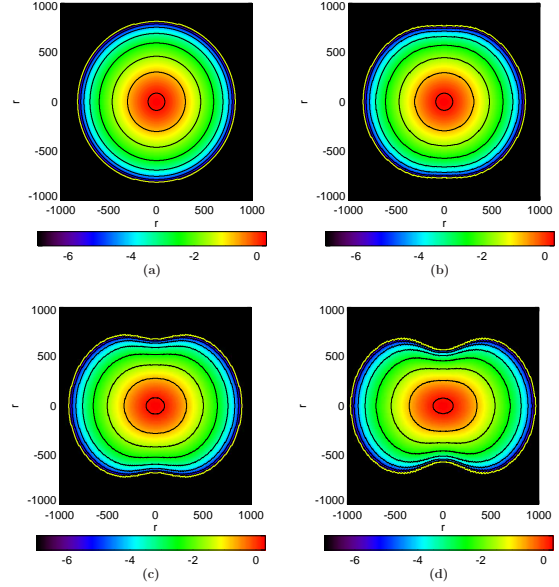}
\caption[Sequence 1: Non-magnetised differentially rotating configurations with changing $\Omega_c$, $A^2=10^5$ fixed]{Sequence of non-magnetised differentially rotating configurations with changing $\Omega_c$. The panels are contour plots of log$\left(\frac{\rho}{\rho_0}\right)$ corresponding to the $\Omega_c$ values (a) 2.028, (b) 12.168, (c) 22.308, (d) 32.448, for which various physical quantities are listed in Table 1. }
\label{fig:rotO}
\end{figure}

Fixing the differential rotation parameter $A$ (see \S 2.2 of \citealt{KEH1}, for
details of the physical meaning of $A$), we vary the central 
angular velocity $\Omega_c$. The stellar density profiles in Fig. 
\ref{fig:rotO} show the development of polar hollows with the increase in 
$\Omega_c$. In fact, the polar radius ($r_p$) decreases, whereas the equatorial
 radius ($r_e$) increases, with the increase of $\Omega_c$, as might be expected 
due to increased centrifugal forces. This results in oblate structures. 
The appearance of polar hollow is due to the differential rotation of the star.
A faster rotation in the inner region disperses the matter to the outer region,
making the central region of the star very oblate. However, the white dwarfs become less oblate in the 
outer region, with the decrease of rotational frequency, hence develops
a polar hollow.
For a KE/GE=0.14, and equatorial angular velocity at the outer surface $\Omega_{eq}=3.089$ 
corresponding to a rotational time period $\sim2$s, the mass is about 
$1.826$ --- an increase of nearly 29\% from the non-rotating case mass 
of 1.416. Table 1 shows the effects of increasing $\Omega_c$
and $\Omega_{eq}$ in the white dwarfs.

\begin{figure}
\centering
\includegraphics[width=\linewidth]{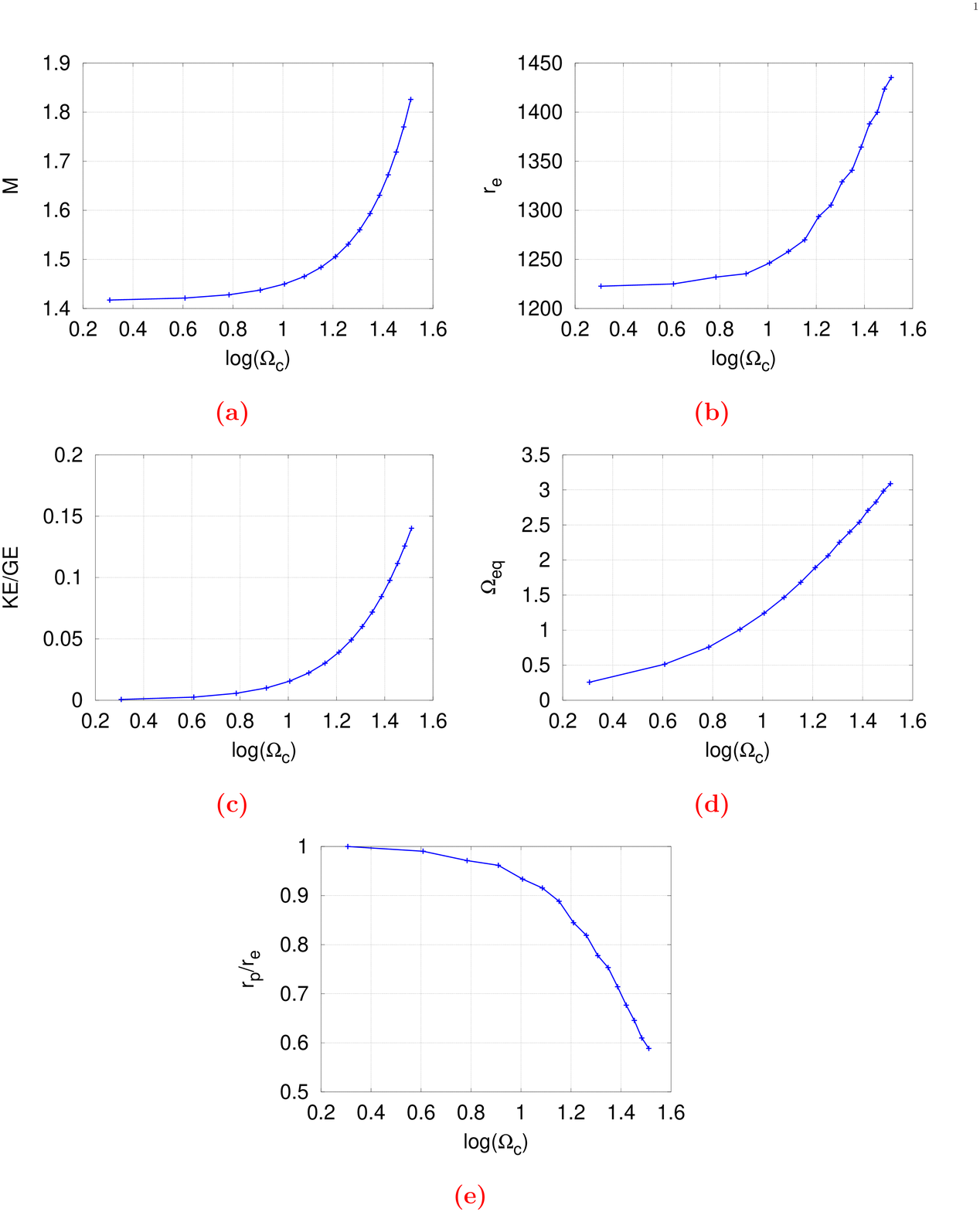}
\caption[Graphs: Sequence 1]{Non-magnetised differentially rotating sequence 
with changing $\Omega_c$. The panels depict 
(a) mass, (b) equatorial radius, (c) KE/GE, (d) equatorial angular velocity, 
(e) $r_p/r_e$, as functions of central rotation rate. }
\label{rotO}
\end{figure}

In Fig. \ref{rotO}, we show the equilibrium sequences
of various stellar quantities.
There appears to be no natural limit, such as the Keplerian limit for uniform 
rotation, for differentially rotating objects with the rotation profile of equation
(\ref{jProfile}). The ratio KE/GE apparently increases without bound. However, instabilities are expected to terminate the sequence at KE/GE $\approx$ 0.27.

\label{shedConf}
For comparison, we note that the uniformly rotating configuration of highest 
$\Omega$, that we were able to compute, has the features: $\Omega_c=5.778$, $M=1.455$, $r_e=1766$, $r_p/r_e=0.672$ and KE/GE=0.0176. This is 
actually quite close to the mass-shedding limit. In fact, naively computing 
$\Omega_{eq}^2r_e/g_e$, where $g_e=GM/r_e^2$, we find it to be $\sim0.95$. 
We were unable to bring {\it XNS} to converge to a solution at higher values of $\Omega_c$ at the central density chosen here.

% float containing table and contour plots
% configs given correspond to XNS omg_c = 1, 6, 11, 16 e-5 
\begin{table}
\centering
\begin{tabular}{|c|c|c|c|c|c|}
\hline 
$\Omega_c$  & $M$ & $r_e$ & $\Omega_{eq}$ & KE/GE & $r_p/r_e$\tabularnewline
\hline 
\hline 
2.028 & 1.417 & 1223 & 0.257 & $6\times10^{-4}$ & 1.000 \tabularnewline
\hline 
12.168 & 1.465 & 1258 & 1.466 & 0.022 & 0.915 \tabularnewline
\hline 
22.308 & 1.593 & 1341 & 2.401 & 0.072 & 0.753 \tabularnewline
\hline 
32.448 & 1.826 & 1435 & 3.089 & 0.140 & 0.588 \tabularnewline
\hline 
\end{tabular}
\caption{Non-magnetised differentially rotating configurations with changing $\Omega_c$.}
\end{table}

% stars - contour plots of log-density

% graphs

% \clearpage			% to enforce float placement

\subsection{Differentially rotating sequence with varying $A^2$}
\begin{figure}
\centering
\includegraphics[width=\linewidth]{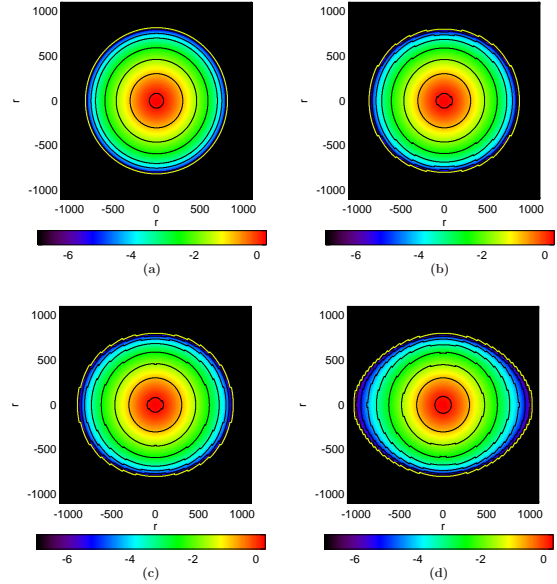}
\caption[Sequence 2: Non-magnetised rotating configurations with changing $A^2$, $\Omega_c=5.778$ fixed]{Sequence of non-magnetised differentially rotating configurations with changing $A^2$ and $\Omega_c=5.778$ fixed. The panels are contour plots of log$\left(\frac{\rho}{\rho_0}\right)$ corresponding to the $A^2$ values (a) 0, (b) $5\times 10^5$, 
(c) $2.5\times 10^6$, (d) $10^7$, for which various physical quantities are listed in the Table 2.}
\label{fig:rotA2}
\end{figure}

\begin{figure}
\centering
\includegraphics[width=\linewidth]{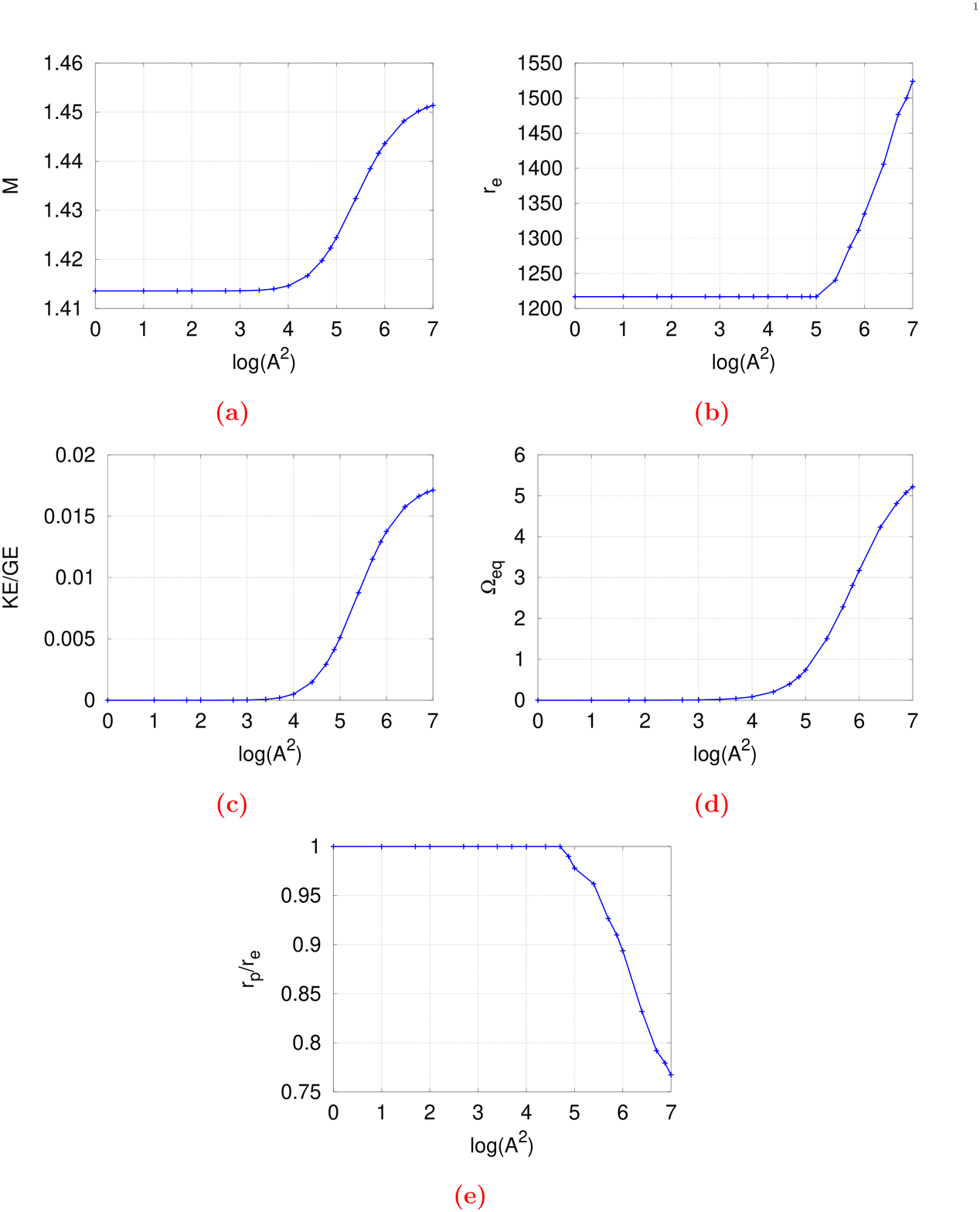}
\caption[Graphs: Sequence 2]{Non-magnetised differentially rotating sequence 
with changing $A^2$, at fixed $\Omega_c=5.778$. The panels depict 
(a) mass, (b) equatorial radius, (c) KE/GE, (d) equatorial angular velocity, 
(e) $r_p/r_e$, as functions of the rotation parameter $A^2$.
}
\label{rotNoBA2}
\end{figure}

% \vspace {-0.1in}
We fix $\Omega_c=5.778$, which is close to the mass-shedding limit for 
the uniformly rotating configuration of the chosen central density. 
Figure \ref{fig:rotA2} and Table 2 show that as $A^2 \to \infty$, we recover 
the uniformly rotating configuration (see \S\ref{shedConf}). We note that the 
sequences of stars in Figs. \ref{fig:rotA2} and \ref{rotNoBA2} show no sign of polar hollows. Our studies indicate that polar hollows develop only when $\Omega_c$ is larger than the angular velocity of the corresponding uniformly rotating model at the mass-shedding limit.

% float containing table and contour plots
\begin{table}
\centering
\begin{tabular}{|c|c|c|c|c|c|}
\hline 
$A^2$  & $M$ & $r_e$ & $\Omega_{eq}$ &  KE/GE & $r_p/r_e$ \tabularnewline
\hline 
\hline 
0 & 1.414 & 1217 & 0 & 0 & 1.000 \tabularnewline
\hline 
$5\times 10^5$ & 1.438 & 1288 & 2.2838 & 0.0115 & 0.927 \tabularnewline
\hline 
$2.5\times 10^{6}$ & 1.448 & 1406 & 4.234 & 0.0158 & 0.832 \tabularnewline
\hline 
$10^{7}$ & 1.451 & 1524 & 5.219 & 0.0171 &  0.767 \tabularnewline
\hline 
\end{tabular}
\caption{Non-magnetised differentially rotating configurations with changing $A^2$, $\Omega_c=5.778$ fixed.}
\end{table}

% stars - contour plots of log-density

% graphs

%\clearpage			% to enforce float placement

%%%%%%%%%%%%%%%%%%%%%%%%%%%%%%%%%%%%%%%%%%%%%%%%%%%%%%%%%%%%%%%%%%%%%%%%%%%%%%%

\section{Magnetised configurations with toroidal field}
We choose a base configuration, by looking for a case where the ratio of magnetic to gravitational 
energies (ME/GE) and KE/GE
are nearly equal, around which to construct the various sequences. The motivation behind this is that such a 
situation is roughly indicative of an equipartition of the effects 
on the white dwarf structure between the rotation and the magnetic field. Our base configuration has the features: 
maximum magnetic field $B_{max}=3.227$, $m=1.4$, $\Omega_c=30.42$, $\Omega_{eq}=0.988$, ME/GE=0.144, KE/GE=0.136, $M=2.586$, and $r_e=2543$. 
%where the number $m$ is the toroidal magnetisation index that determines the form of the barotropic magnetic field profile (see, e.g., \citealt{xns1}).

Stability criteria for magnetised and rotating stars in GR are not well known. This is an important question that we plan to take up in a later work. However, based on the prescription for twisted-torus geometries,
\cite{ciolfi13} showed that compact stars with highly
toroidally dominated magnetic field could be stable and realistic. For the sequences that follow, we choose to keep ME/GE below 0.2, purely as a tentative criterion. It is, however, known that when the magnetic field exceeds $\approx 5$, the EoS is strictly modified due to the effects of Landau quantisation (\citealt{prd12}), and hence we do not allow the field to go beyond $5$ in 
magnitude, as we plan to stick with $\Gamma=4/3$ throughout.

\subsection{No rotation}
\begin{figure}
\centering
\includegraphics[width=\linewidth]{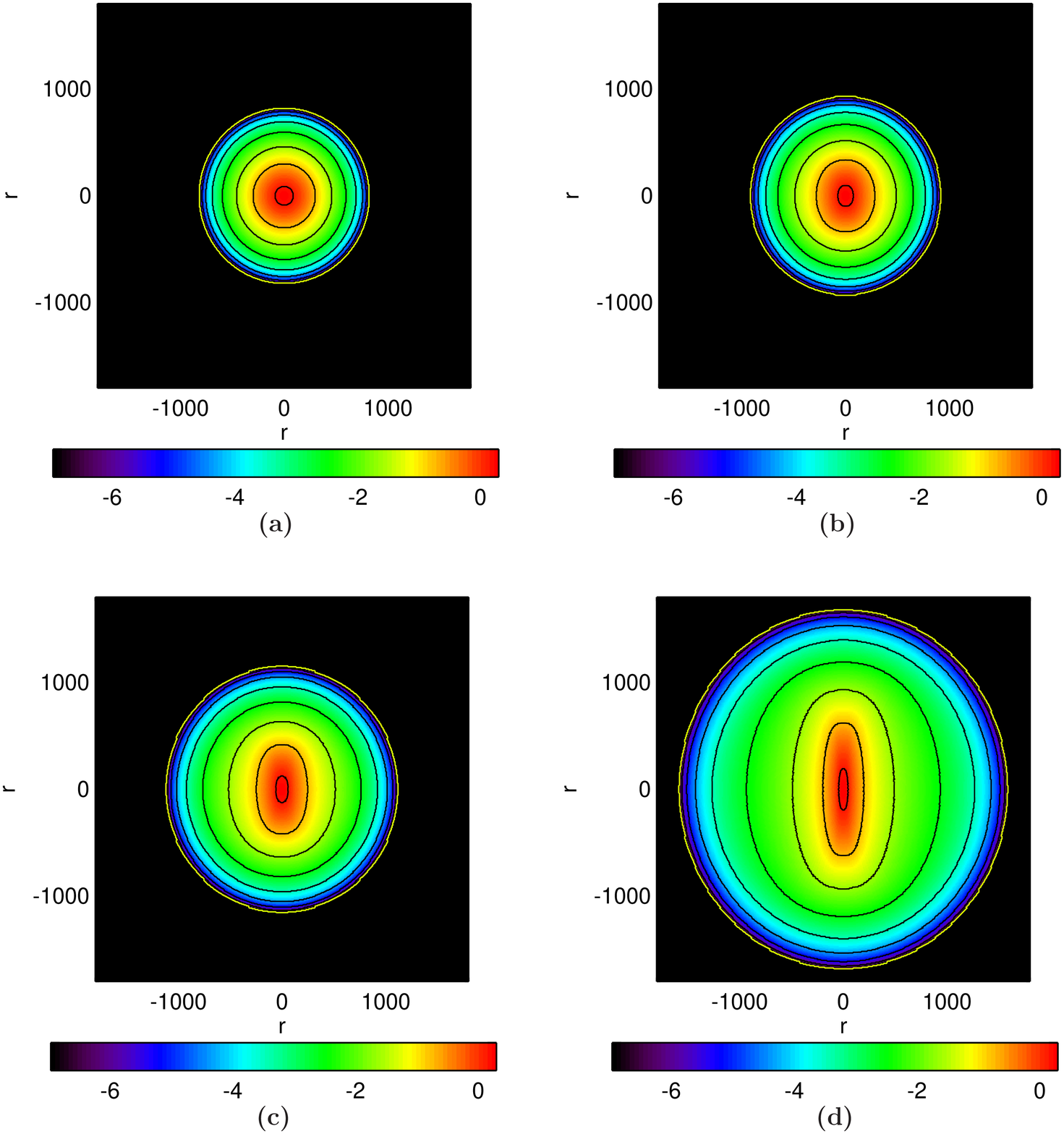}
\caption[Sequence 3: Non-rotating configurations with purely toroidal magnetic field, with changing $B_{max}$ ]{Sequence of non-rotating 
configurations with a purely toroidal magnetic field, with changing $B_{max}$. 
The panels are contour
 plots of log$\left(\frac{\rho}{\rho_0}\right)$ corresponding to the $B_{max}$ values (a) 0, (b) $2.028$, (c) $2.822$, 
(d) $3.405$. The corresponding physical quantities are listed 
in Table 3. }
\label{fig:torNoRot}
\end{figure}

The physics of magnetised white dwarfs is complicated by the fact that the 
pressure becomes anisotropic due to the inclusion of a magnetic field pressure 
and a magnetic tension (\citealt{jcap14}). The magnetic tension for a toroidal 
field tends to cause the star to become prolate, stretching it along the polar direction. 
This is indeed seen from the density isocontours in Fig. \ref{fig:torNoRot} and Table 3. 
We also observe that the magnetic field does not influence the structure of 
the star significantly until it attains a magnitude of about $10^{14}$G. This 
can be understood physically from the fact that the magnetic pressure term, 
$B^2/8\pi$, becomes comparable to the matter pressure in a white dwarf of 
central density $\sim 10^{10}$g/cc, with the EoS given by equation (\ref{EoS}), 
at this magnitude of field only. 

\begin{figure}
\centering
\includegraphics[width=\linewidth]{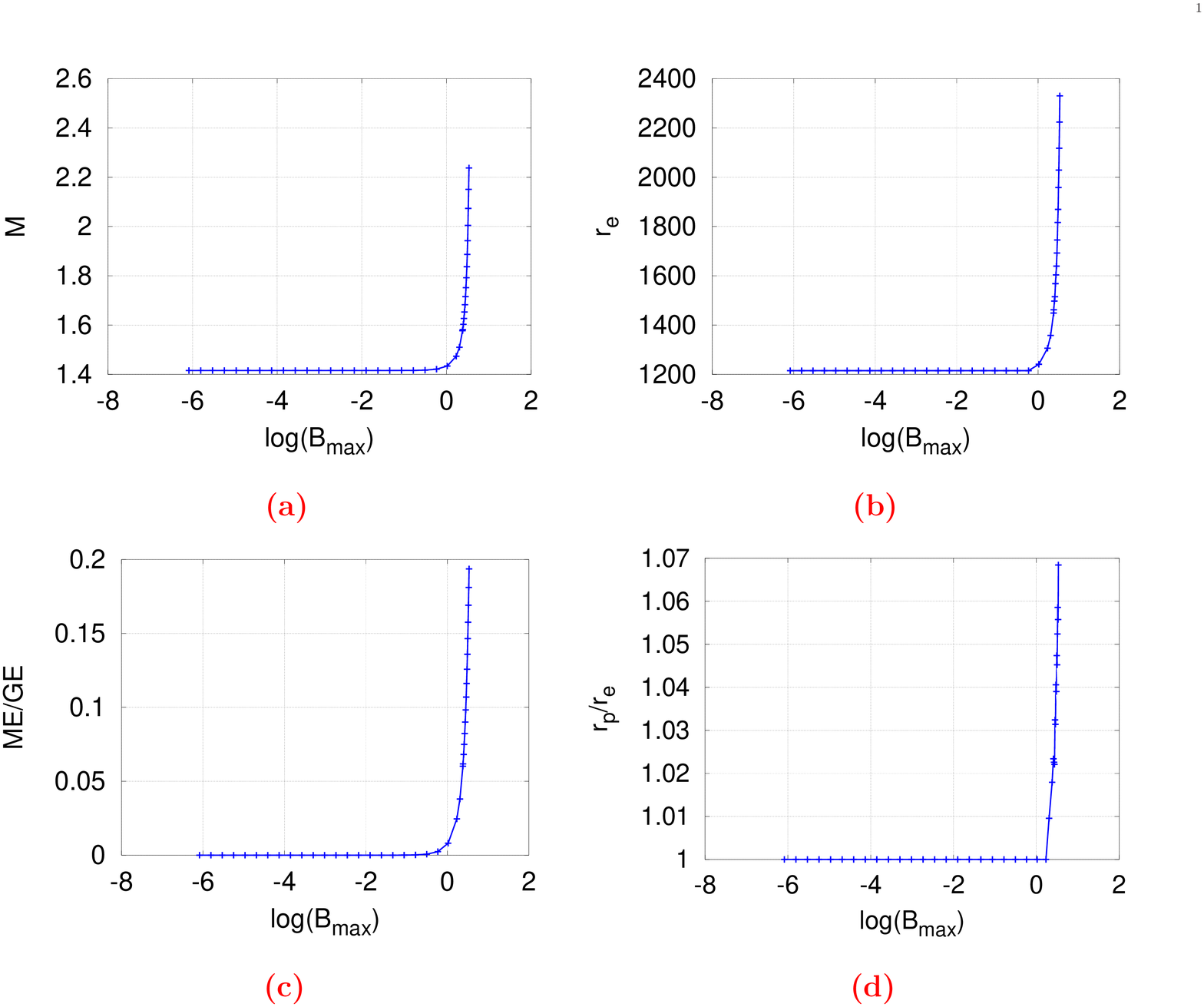}
\caption[Graphs: Sequence 3]{Non-rotating sequence with changing $B_{max}$ of toroidal magnetic field. The panels depict  
(a) mass, (b) equatorial radius, (c) ME/GE, (d) $r_p/r_e$,  
as functions of the maximum magnetic field. }
\label{torNoRot}
\end{figure}

Figure \ref{torNoRot} shows that with increasing field, both $r_e$ and $r_p$ 
increase, but $r_p$ increases more than $r_e$ leading to prolateness. However, 
the apparent structure is still nearly spherical, as $r_p/r_e$ is atmost 
1.07. We see that for a $B_{max}$ of about $3.4$, the mass 
increases to $2.238$, by about $58\%$ from 1.416 
in the non-magnetised case. However, the value of $r_e$, 2330, is nearly double that in the non-magnetised case, so that the mean density in comparison goes 
down by a factor of nearly 10. This also aligns well with the fact that the density isocontours show a decrease in the degree of central condensation.

Our computations reveal that ME/GE seems to increase without bound, with the 
code still capable of converging to an equilibrium solution for values of 
ME/GE even higher than $0.3$, for example. This makes the need for a stability analysis
 all the more significant, and we also plan to tackle this problem in a
future work. 

\begin{table}
\centering
\begin{tabular}{|c|c|c|c|c|}
\hline 
$B_{max}$  & $M$ & $r_e$ & ME/GE & $r_p/r_e$\tabularnewline
\hline 
\hline 
0 & 1.416 & 1215 & 0 & 1 \tabularnewline
\hline 
2.028 & 1.510 & 1358 & 0.038 & 1.010 \tabularnewline
\hline 
2.822 & 1.715 & 1639 & 0.098 & 1.032 \tabularnewline
\hline 
3.405 & 2.238 & 2330 & 0.194 & 1.068 \tabularnewline
\hline 
\end{tabular}
\caption{Non-rotating configurations with purely toroidal magnetic field, with changing $B_{max}$.}
\end{table}

% stars - contour plots of log-density

% graphs

\subsection{Uniform rotation}
We seek to construct a sequence of varying $\Omega_c$, keeping the magnetic field fixed. In practice, although we fix the two constants that determine the magnetic field profile, there is a small change in the field with the 
change of $\Omega_c$. This is due to the computation of a self-consistent solution. The effect is very small, the fractional change being $\sim 10^{-3}$. 
The initial fixed $B_{max}$ is $3.092$, and becomes $3.094$ when $\Omega_c$ is increased from 0 to 3.4. 

\begin{figure}
\centering
\includegraphics[width=\linewidth]{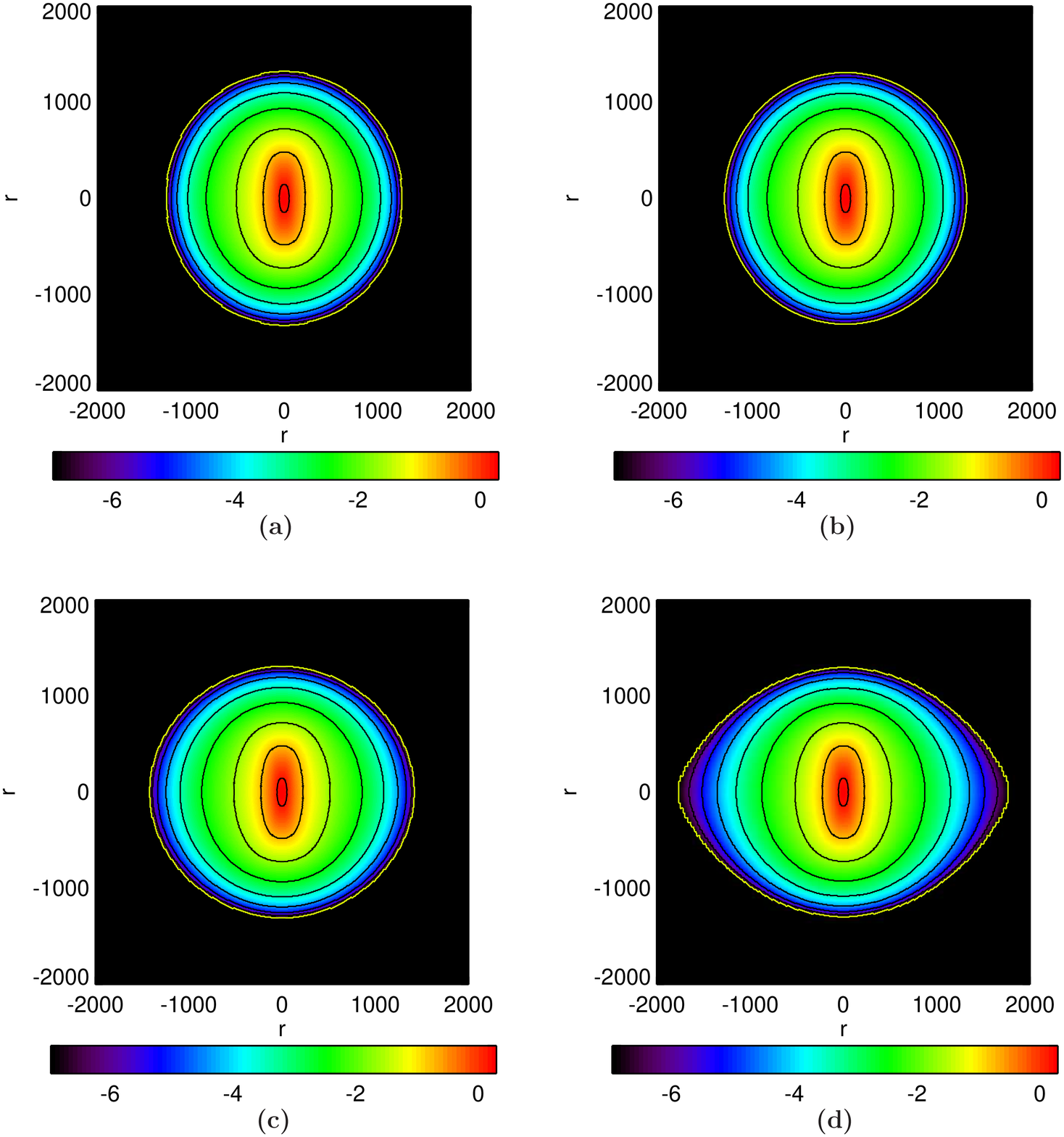}
\caption[Sequence 4: Uniformly rotating configurations with purely toroidal magnetic field, with changing $\Omega_c$, $B_{max}=$, and $m=1.4$ fixed]{Sequence of uniformly rotating configurations with a purely toroidal 
magnetic field with changing $\Omega_c$ at a fixed magnetic field having 
$B_{max}=3.092$. The panels are contour plots of log$\left(\frac{\rho}{\rho_0}\right)$ corresponding to the $\Omega_c$ values (a) 0, (b) 1.358, (c) 2.620, 
(d) 3.407. The corresponding physical quantities are listed in Table 4. 
}
\label{fig:torUniRot}
\end{figure}

\begin{figure}
\centering
\includegraphics[width=\linewidth]{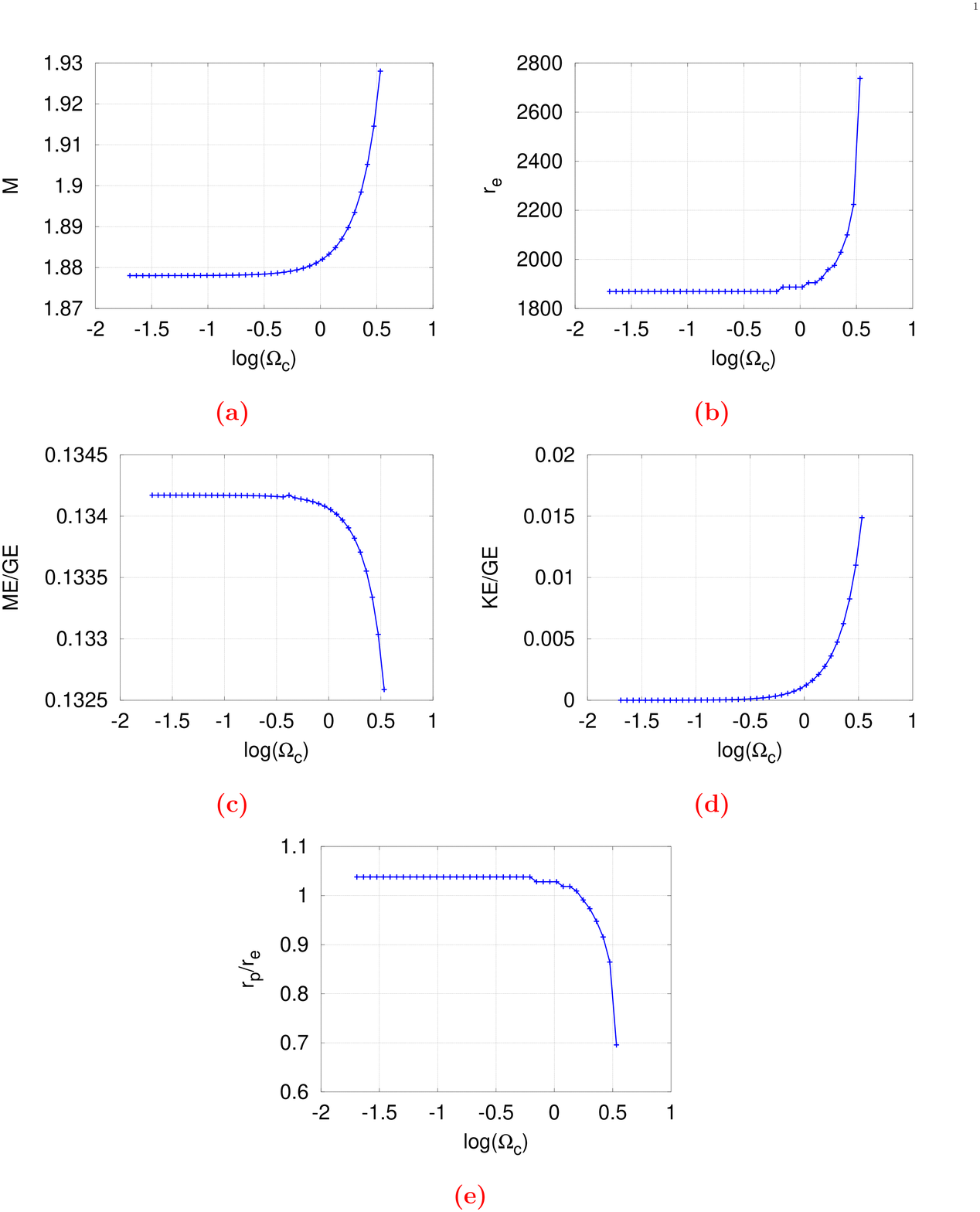}
\caption[Graphs : Sequence 4]{Uniformly rotating sequence with changing 
$\Omega_c$ at fixed toroidal $B_{max}=3.092$.
The panels depict 
(a) mass, (b) equatorial radius, (c) ME/GE, (d) KE/GE, (e) $r_p/r_e$,
as functions of the central rotation rate.
  }
\label{torUniRot}
\end{figure}

Only in the far outer layers, where the centrifugal term (proportional to $r$) 
becomes large, the density isocontours become oblate ---  the deformation typical of 
axisymmetric rotation. The interior layers show mainly a magnetic field 
induced deformation, which is prolate for toroidal fields, as shown in
Fig. \ref{fig:torUniRot}. ME/GE remains nearly constant at 0.133 as $\Omega_c$ is changed, 
and KE/GE increases from 0 to about 0.015, as shown in Fig. \ref{torUniRot}. 

The ratio $\Omega_{eq}^2r_e/g_e$ comes out to be $\sim 0.93$ for $\Omega_c=3.407$. The saturation of the effect of uniform rotation can be seen in the drastic decrease of $r_p/r_e$, from 1.038 to 0.696, approaching the value at the mass-shedding limit for the non-magnetised case, as given in Table 4 and shown in Fig. \ref{torUniRot}. The mass increase due to the effect of rotation on the initial non-rotating magnetised case is about $0.05$, which appears to be the same as for the non-magnetised case, given by Table 2. However, note that due to the problem of resolution, Figs. \ref{torUniRot}(b) 
and (e) do not appear to be smooth (show discrete behaviour, which is not physical).
At higher resolutions, much larger runtimes
are taken to compute a configuration, without revealing any meaningful physical change in the results. However
for the primary understanding of the problem, our
resolution suffices and the trends obtained are clear.

%\newpage

%\clearpage			% to enforce float placement

% float containing table and contour plots
% configs given correspond to XNS omg_c = 0, 6.697e-6, 1.2917e-5, ,1.68e-5
\begin{table}
\centering
\begin{tabular}{|c|c|c|c|c|c|}
\hline 
$\Omega_c$  & $M$ & $r_e$ & KE/GE & ME/GE & $r_p/r_e$\tabularnewline
\hline 
\hline 
0 & 1.878 & 1869 & 0 & 0.1342 & 1.038 \tabularnewline
\hline 
1.358 & 1.885 & 1905 & 0.002 & 0.1339 & 1.019 \tabularnewline
\hline 
2.620 & 1.905 & 2100 & 0.008 & 0.1333 & 0.916 \tabularnewline
\hline 
3.407 & 1.928 & 2738 & 0.015 & 0.1326 & 0.696 \tabularnewline
\hline 
\end{tabular}
\caption{Uniformly rotating configurations with purely toroidal magnetic field, with changing $\Omega_c$ and 
$B_{max}=3.092$ fixed.}
\end{table}

% stars - contour plots of log-density

% graphs

%\clearpage			% to enforce float placement

\subsection{Differential rotation}
Physically, differential rotation interacts with magnetic fields through 
rotationally induced currents. Hence, there is a rich range of phenomena 
associated with rotating magnetised systems, such as magnetic braking and 
relativistic jets from accreting systems etc. \citet{rotReview} give a 
wide overview of the work that has been done on these effects.

\subsubsection{Sequence with fixed $B_{max}$ constructed by varying $\Omega_c$}
\begin{figure}
\centering
\includegraphics[width=\linewidth]{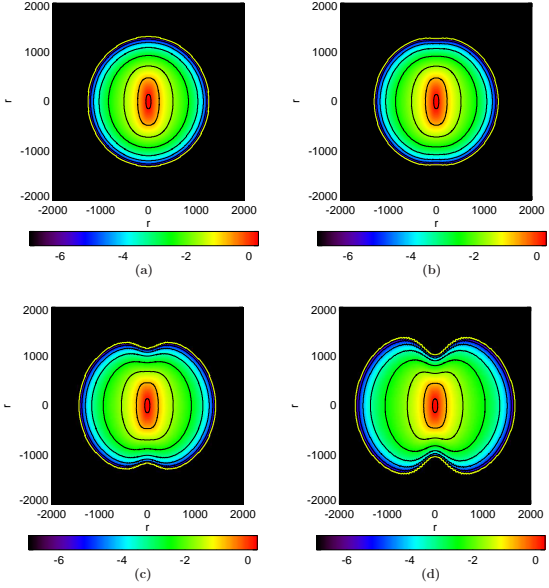}
\caption[Sequence 5: Differentially rotating configurations with purely toroidal magnetic field, with changing $\Omega_c$ and $B_{max}\approx 3.1$ and $m=1.4$ fixed]{Sequence of differentially rotating 
configurations with a purely toroidal magnetic field with changing $\Omega_c$ and
$B_{max}\approx 3.1$ fixed. The panels are contour plots of 
log$\left(\frac{\rho}{\rho_0}\right)$ corresponding to the $\Omega_c$ values 
(a) 0.003, (b) 8.112, (c) 18.252, (d) 28.392. The corresponding physical 
quantities are listed in Table 5. 
 }
\label{fig:torO}
\end{figure}

\begin{figure}
\centering
\includegraphics[width=\linewidth]{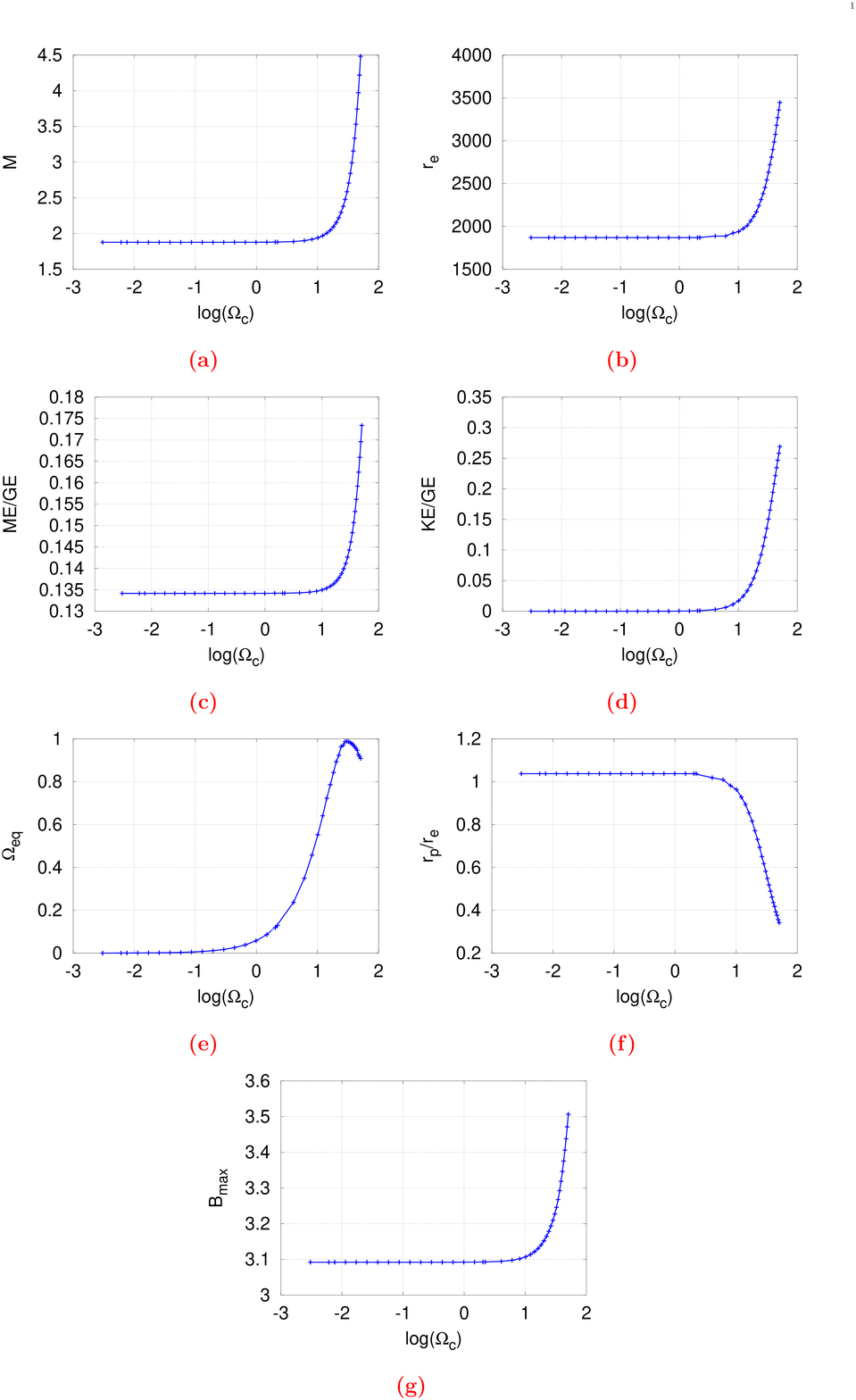}
\caption[Graphs: Sequence 5]{Differentially rotating sequence with changing 
$\Omega_c$ at fixed toroidal $B_{max}\approx 3.1$. 
The panels depict 
(a) mass, (b) equatorial radius, (c) ME/GE, (d) KE/GE, (e) equatorial angular 
velocity, (f) $r_p/r_e$, (g) $B_{max}$,
as functions of the central rotation rate. 
  }
\label{torO}
\end{figure}

As before, the magnetic profile is fixed with $m=1.4$, but the magnitude of $B_{max}$ slightly changes with $\Omega_c$. 
As Table 5 and Fig. \ref{fig:torO} show, for the range of results that we present based on our 
tentative stability criteria, $B_{max}$ does not change much. However, the 
Fig. \ref{torO} shows that $B_{max}$, $M$, $r_e$, KE/GE, ME/GE, all have quite a 
sharp increasing trend near the highest value of $\Omega_c=50$ in our sequence. 
This might be indicative of some limiting behaviour, and needs to be investigated analytically. 

We note from Fig. \ref{fig:torO} that the central regions of the stars are prolate, 
with strong polar hollows at higher $\Omega_c$. Along the sequence, $r_p/r_e$ 
decreases, falling sharply near $\Omega_c=50$. The outer regions are relatively oblate with increasing $\Omega_c$. 
The polar hollow appears to be sharper as compared to a case with no magnetic field but similar 
(but slightly smaller) rotation rate. The toroidal 
field peaks slightly away from the centre of the star, and the distortion of the outer density isocontours to a prolate nature is pronounced at this radius. This could explain why the hollows, which are induced primarily by differential rotation, are accentuated by the toroidal field.

Interestingly, $\Omega_{eq}$ increases at first, peaks and starts decreasing. 
The mass increases by $\sim 32\%$, 
from $1.878$ for the non-rotating configuration to 2.478 
for a KE/GE value of 0.121, and equatorial rotation period of about 6.5 seconds, as shown in Table 5. The 
equatorial radius increases to 2454, marginally more than the value for 
a field of similar magnitude but without rotation.

% float containing table and contour plots
% configs given correspond to XNS omg_c = 1.5e-8, 4e-5, 9e-5, 14e-5
\begin{table}
\centering
\large
\resizebox{\linewidth}{!}{
\begin{tabular}{|c|c|c|c|c|c|c|c|}
\hline 
$\Omega_c$  & $M$ & $r_e$ & $\Omega_{eq}$ & $B_{max} $ & KE/GE & ME/GE & $r_p/r_e$\tabularnewline
\hline 
\hline 
0.003 & 1.878 & 1869 & 0.0002 & 3.0921 & $1.5{\times}10^{-9}$ & 0.134 & 1.038 \tabularnewline
\hline 
8.112 & 1.918 & 1922 & 0.458 & 3.1016 & 0.011 & 0.135 & 0.982 \tabularnewline
\hline 
18.252 & 2.097 & 2118 & 0.843 & 3.1407 & 0.054 & 0.137 & 0.816 \tabularnewline
\hline 
28.392 & 2.478 & 2454 & 0.988 & 3.2098 & 0.121 & 0.143 & 0.617 \tabularnewline
\hline 
\end{tabular}
}
\caption{Differentially rotating configurations with purely toroidal magnetic field with changing $\Omega_c$
and $B_{max}\approx 3.1$ fixed.}
\end{table}

% stars - contour plots of log-density

% graphs

\subsubsection{Sequence with fixed $\Omega_c$ constructed by varying $B_{max}$}
For a $B_{max}$ of $3.584$, the mass increases to about 3.159, by nearly 78\% from the mass in the non-magnetised case. The radius behaves in nearly the same way as for the purely magnetised case, and $r_e$ becomes about 3322, more than double that of the non-magnetised configuration. The surface $\Omega_{eq}$ 
reduces from 2.99 to 0.593, as $B_{max}$ increases. This could be understood as the increase of $r_e$ increases 
centrifugal forces at the equator for a given $\Omega_{eq}$, which enforces decreased $\Omega_{eq}$ in order 
to obtain equilibrium solutions. See Table 6 for details.

\begin{figure}
\centering
\includegraphics[width=\linewidth]{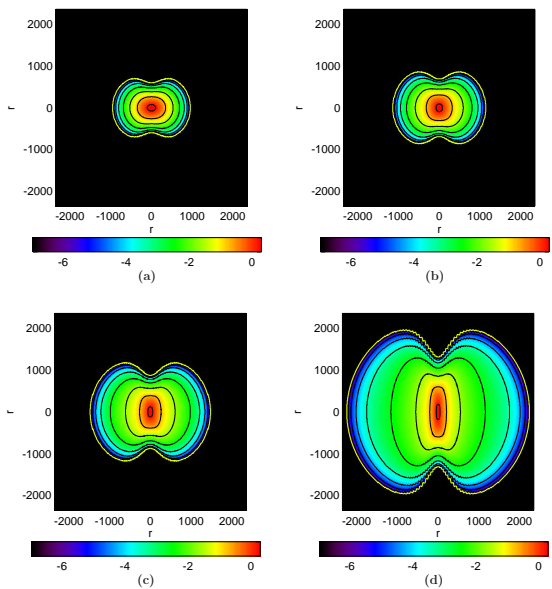}
\caption[Sequence 6: Differentially rotating configurations with purely toroidal 
magnetic field, with changing $B_{max}$,$m=1.4$ and $\Omega_c=30.42$ fixed]
{Sequence of differentially rotating configurations with a purely toroidal magnetic 
field, with changing $B_{max}$ and fixed $\Omega_c=30.42$.
The panels are contour plots of 
log$\left(\frac{\rho}{\rho_0}\right)$ corresponding to the $B_{max}$ values (a) 0, 
(b) $2.299$, (c) $2.996$, (d) $3.584$. 
The corresponding physical quantities are listed in Table 6.  }
\label{fig:torB}
\end{figure}

We note again that the polar concavities, which develop primarily due to 
differential rotation, are accentuated by the toroidal field. Figure 
\ref{fig:torB} clearly shows the severe polar denting with the increase in $B_{max}$. 

\begin{figure}
\centering
\includegraphics[width=\linewidth]{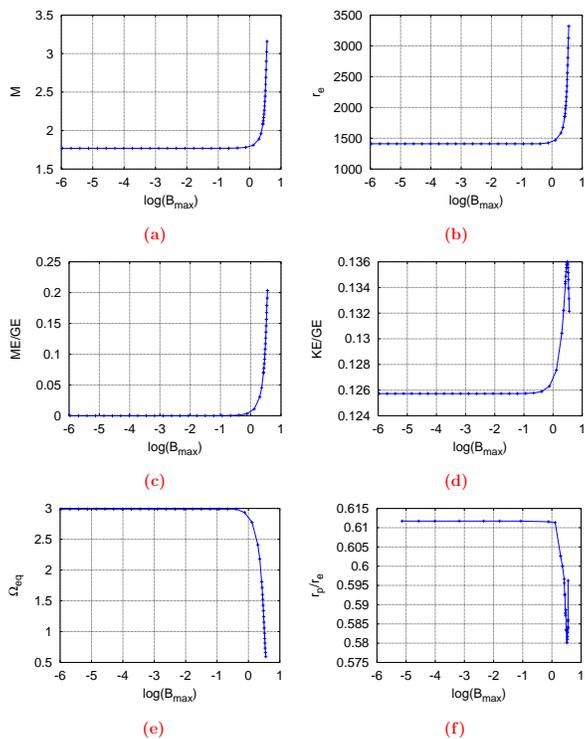}
\caption[Graphs: Sequence 6]{Differentially rotating sequence with changing toroidal $B_{max}$ 
and fixed $\Omega_c=30.42$. The panels depict 
 (a) mass, (b) equatorial radius, (c) ME/GE, (d) KE/GE, (e) equatorial angular velocity,
(f) $r_p/r_e$,  
as functions of the maximum magnetic field.
}
\label{torB}
\end{figure}

An interesting feature seen in Fig. \ref{torB} is that with the increase of
$B_{max}$, KE/GE increases initially, attains a peak value of 0.136, and then 
starts decreasing. The value of $r_p/r_e$ too follows the corresponding pattern with a dip,
when the largest KE/GE corresponds to the smallest $r_p/r_e$. We 
speculate that this may indicate the onset of some instability, and a possible 
upper limit to the mass in this sequence as well, and plan to follow it up in 
a future work.

% float containing table and contour plots
% configs given correspond to XNS bcoef = 0, 35, 74, 150
\begin{table}
\centering
\resizebox{\linewidth}{!}{
\begin{tabular}{|c|c|c|c|c|c|c|}
\hline 
$B_{max}$  & $M$ & $r_e$ & $\Omega_{eq}$ & KE/GE & ME/GE & $r_p/r_e$\tabularnewline
\hline 
\hline 
0 & 1.769 & 1410 & 2.990 & 0.126 & 0 & 0.613 \tabularnewline
\hline 
2.299 & 1.959 & 1676 & 2.180 & 0.132 & 0.046 & 0.603 \tabularnewline
\hline 
2.996 & 2.318 & 2171 & 1.339 & 0.136 & 0.108 & 0.583 \tabularnewline
\hline 
3.584 & 3.159 & 3322 & 0.593 & 0.132 & 0.203 & 0.584 \tabularnewline
\hline 
\end{tabular}}
\caption{Differentially rotating configurations with purely toroidal magnetic 
field, with changing $B_{max}$ and $\Omega_c=30.42$ fixed.}
\end{table}

% stars - contour plots of log-density

% graphs

%\clearpage			% to enforce float placement

%%%%%%%%%%%%%%%%%%%%%%%%%%%%%%%%%%%%%%%%%%%%%%%%%%%%%%%%%%%%%%%%%%%%%%%%%%%%%%%

\section{Magnetised configurations with poloidal field}
\cite{Fer37} showed that for a steady state, in a rotating star having a poloidal magnetic field, each field line must rotate uniformly. This result, called the `isorotation law', dictates that the most general velocity field for 
a rotating star can only have uniform rotation of each poloidal flux loop. This is a strong restriction, and makes the computation of models with poloidal fields more difficult.

Poloidal fields are seen to result in highly centrally condensed and compact configurations. Both $r_e$ and $r_p$ tend to decrease with the increase of $B_{max}$, with the development of oblateness (equatorial bulge) due to a faster decrease of $r_p$ than $r_e$. In this respect, poloidal magnetic fields have an effect similar to rotation.

Our sequences having a purely poloidal magnetic field have been constructed around the base configuration 
(reasons for this choice are discussed in the beginning of \S 5) with the features: $B_{max}=4.575$, $\Omega_c=30.42$, $\Omega_{eq}=4.950$, ME/GE=0.0773, KE/GE=0.106, $M=1.983$, and $r_e=1054$. 

\subsection{No rotation}
\begin{figure}
\centering
\includegraphics[width=\linewidth]{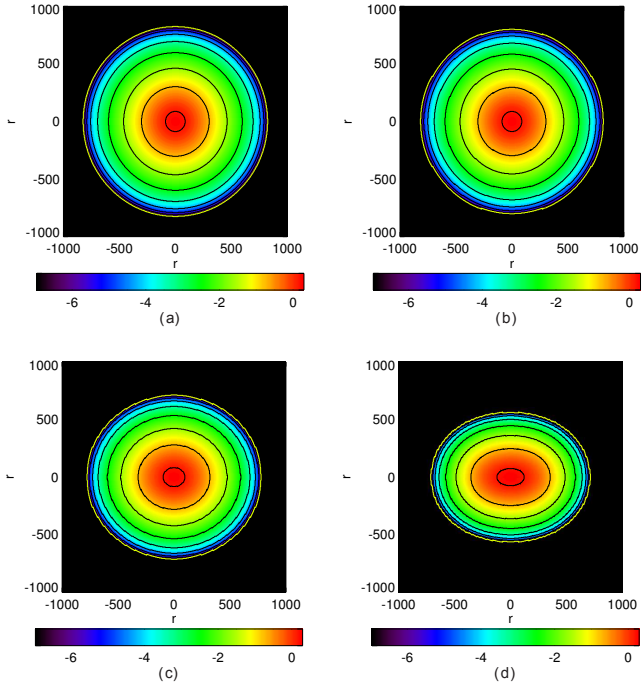}
\caption[Sequence 7: Non-rotating configurations with purely poloidal magnetic field, 
with changing $B_{max}$ but fixed magnetic field profile]{Sequence of non-rotating 
configurations with a purely poloidal magnetic field with changing $B_{max}$.
The panels are contour plots of log$\left(\frac{\rho}{\rho_0}\right)$ 
corresponding to the $B_{max}$ values (a) 0, (b) $0.846$, (c) $2.260$, 
(d) $3.942$. The corresponding physical quantities are listed in Table 7.  }
\label{fig:polNoRot}
\end{figure}

\begin{figure}
\centering
\includegraphics[width=\linewidth]{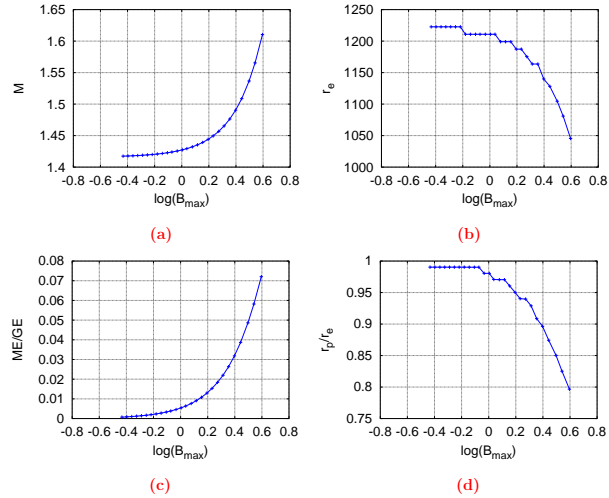}
\caption[Graphs: Sequence 7]{Non-rotating sequence with changing $B_{max}$ of poloidal 
magnetic field. The panels depict 
(a) mass, (b) equatorial radius, (c) ME/GE, (d) $r_p/r_e$,
as functions of the maximum magnetic field. 
  }
\label{polNoRot}
\end{figure}

Figures \ref{fig:polNoRot} and \ref{polNoRot} show how the stars become progressively 
more oblate with increasing $B_{max}$. ME/GE is lower than that for a toroidal field of 
similar magnitude. The mass increases from $1.416$, in the non-magnetised case, to 
about $1.610$ with a $B_{max}$ of $3.942$ --- an increase of 
$\sim 14\%$, as given in Table 7. 
However, like Figs. \ref{torUniRot}(b) and (e), here also Figs. \ref{polNoRot}(b) and (d) exhibit 
the resolution problem (showing discrete behaviour, which is not physical), 
but clearly illustrate the trend of the quantity along the sequence.

It is in general observed that the poloidal field 
contributes much less to mass increase than a toroidal field. However, it causes decrease in the size of the star -- $r_e$ decreases from 1215, in the 
non-magnetised case, to about 1000 for $B_{max}=4$. This results in more centrally condensed (having higher mean density) and compact configurations. The polar radius also decreases rapidly, causing oblate structures with $r_p/r_e$ as small as 0.79.

The contrasting effects of poloidal and toroidal fields - poloidal causing greater central condensation and smaller radii, and toroidal causing reduced mean densities and larger radii - are entirely due to the difference in the field geometries. The geometry decides the direction and action of Lorentz forces due the magnetic field (\citealt{Cardall2001}).
% float containing table and contour plots first
% values correspond to XNS kbpol - 0, 2.5e-3, 6.24e-3, 9.44e-3
\begin{table}
\centering
\begin{tabular}{|c|c|c|c|c|}
\hline 
$B_{max}$  & $M$ & $r_e$ & ME/GE & $r_p/r_e$\tabularnewline
\hline 
\hline 
0 & 1.416 & 1215 & 0 & 1 \tabularnewline
\hline 
0.846 & 1.424 & 1210 & 0.0046 & 0.99 \tabularnewline
\hline 
2.260 & 1.476 & 1163 & 0.0264 & 0.91 \tabularnewline
\hline 
3.942 & 1.610 & 1045 & 0.0721 & 0.79 \tabularnewline
\hline 
\end{tabular}
\caption{Non-rotating configurations with purely poloidal magnetic field with changing 
$B_{max}$.}
\end{table}

% stars - contour plots of log-density

% graphs

\subsection{Uniform rotation}
We notice in this case from Table 8 that including rotation in the model with a poloidal 
field seems to cause an abrupt decrease of mass initially, to $1.5$, from $1.58$. At higher rotation rates, it then rises marginally above the mass for the non-rotating star, to $1.62$. However, the sequence numerically terminates 
at similar ME/GE 
and mass as the magnetised non-rotating configuration, so that we are unable to pursue 
the change of $\Omega_c$ furthermore. The magnitude of the field increases from $3.42$ to $3.65$, even though the parameters determining the field are held fixed, indicating that the self-consistent solution requires 
non-negligible changes in the magnetic field for small changes (or no change) in the rotation profile.

\begin{figure}
\centering
\includegraphics[width=\linewidth]{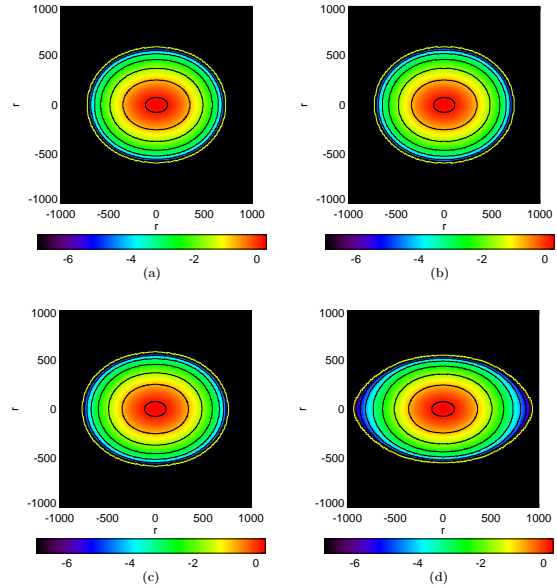}
\caption[Sequence 8: Uniformly rotating configurations with a purely poloidal magnetic field, with changing $\Omega_c$; $B_{max}=3.42$ and field profile fixed]{Sequence of uniformly rotating configurations with a purely poloidal magnetic field, with changing $\Omega_c$ at a fixed magnetic field having $B_{max}=3.42$. The panels are 
contour plots of log$\left(\frac{\rho}{\rho_0}\right)$ corresponding to the $\Omega_c$ 
values (a) 0, (b) 2.432, (c) 5.159, (d) 8.518. The corresponding physical quantities are 
listed in Table 8.  }
\label{fig:polUniRot}
\end{figure}

\begin{figure}
\centering
\includegraphics[width=\linewidth]{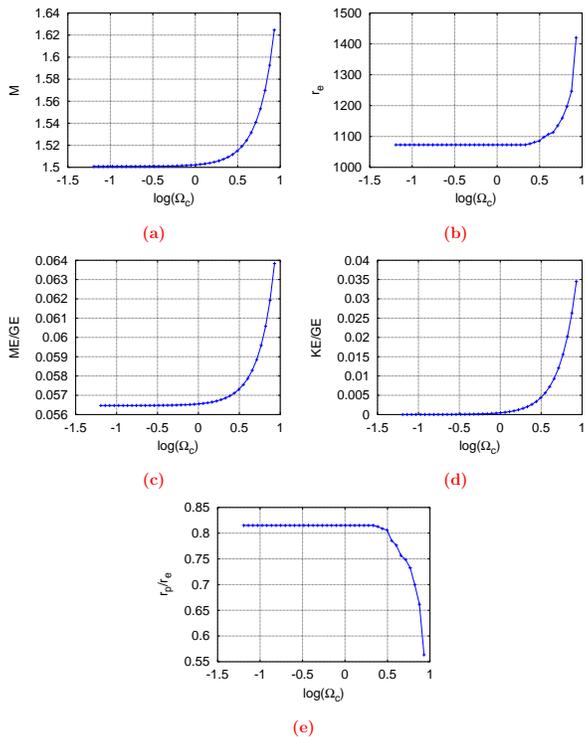}
\caption[Graphs: Sequence 8]{Uniformly rotating sequence with changing $\Omega_c$ at
fixed poloidal $B_{max}=3.42$. The panels depict 
 (a) mass, (b) equatorial radius, (c) ME/GE, (d) KE/GE, (e) $r_p/r_e$,
as functions of the central rotation rate.  }
\label{polUniRot}
\end{figure}

Moreover, the $r_e$ increases, changing from 1073 in the non-rotating case to 1420, for a rotation period of 0.74 seconds, as shown in Figs. \ref{fig:polUniRot} and \ref{polUniRot}. 
This must mean that rotation has a stronger effect than the magnetic field, since the latter tends to cause a reduction 
of radius. The radius ratio decreases much more than its value of 0.67 (not shown in any sequence) and 0.79, in the case of pure rotation (not shown in any sequence)
and pure magnetic field respectively, to 0.56. This is because the poloidal field and rotation both reinforce each other in making the star oblate. 

% float containing table and contour plots
% corresponding to XNS omega_c = 0, 1.199e-5, 2.5437e-5, 4.2e-5
\begin{table}
\centering
\begin{tabular}{|c|c|c|c|c|c|}
\hline 
$\Omega_c$  & $M$ & $r_e$ & KE/GE & ME/GE & $r_p/r_e$\tabularnewline
\hline 
\hline 
0 & 1.582 & 1073 & 0 & 0.064 & 0.803 \tabularnewline
\hline 
2.432 & 1.509 & 1085 & 0.003 & 0.057 & 0.806 \tabularnewline
\hline 
5.159 & 1.541 & 1085 & 0.012 & 0.059 & 0.749 \tabularnewline
\hline 
8.518 & 1.625 & 1420 & 0.035 & 0.064 & 0.563 \tabularnewline
\hline 
\end{tabular}
\caption{Uniformly rotating configurations with a purely poloidal magnetic field with 
changing $\Omega_c$ and $B_{max}=3.42$ fixed.}
\end{table}

% stars - contour plots of log-density

% graphs

%\clearpage	% clear the floats

\subsection{Differential rotation}
In general, the allowed $\Omega_{eq}$ for rotating configurations with poloidal fields is much higher than in the similar cases with toroidal fields. This may be due to the decreased $r_e$ in the former cases, 
which allows for faster rotation at the equator, because centrifugal effects decrease when radius decreases 
for a given angular velocity. 

\subsubsection{Sequence with fixed poloidal $B_{max}$ constructed by varying $\Omega_c$}
\begin{figure}
\centering
\includegraphics[width=\linewidth]{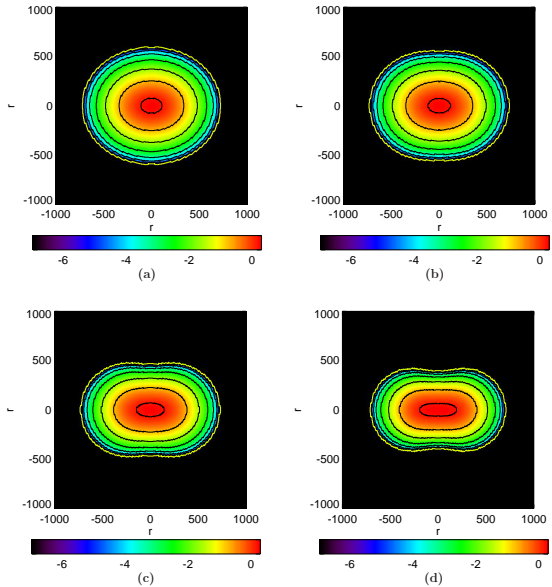}
\caption[Sequence 9: Differentially rotating configurations with purely poloidal magnetic 
field, with changing $\Omega_c$; $B_{max}=3.1$ and field profile fixed]
{Sequence of differentially rotating configurations with a purely poloidal magnetic field
 with changing $\Omega_c$ and $B_{max}=3.1$ fixed.
The panels are contour plots of log$\left(\frac{\rho}{\rho_0}\right)$ 
corresponding to the $\Omega_c$ values (a) 2.028, (b) 12.168, (c) 24.336, (d) 32.448. 
The corresponding physical quantities are listed in Table 9.  }
\label{fig:polO}
\end{figure}

\begin{figure}
\centering
\includegraphics[width=\linewidth]{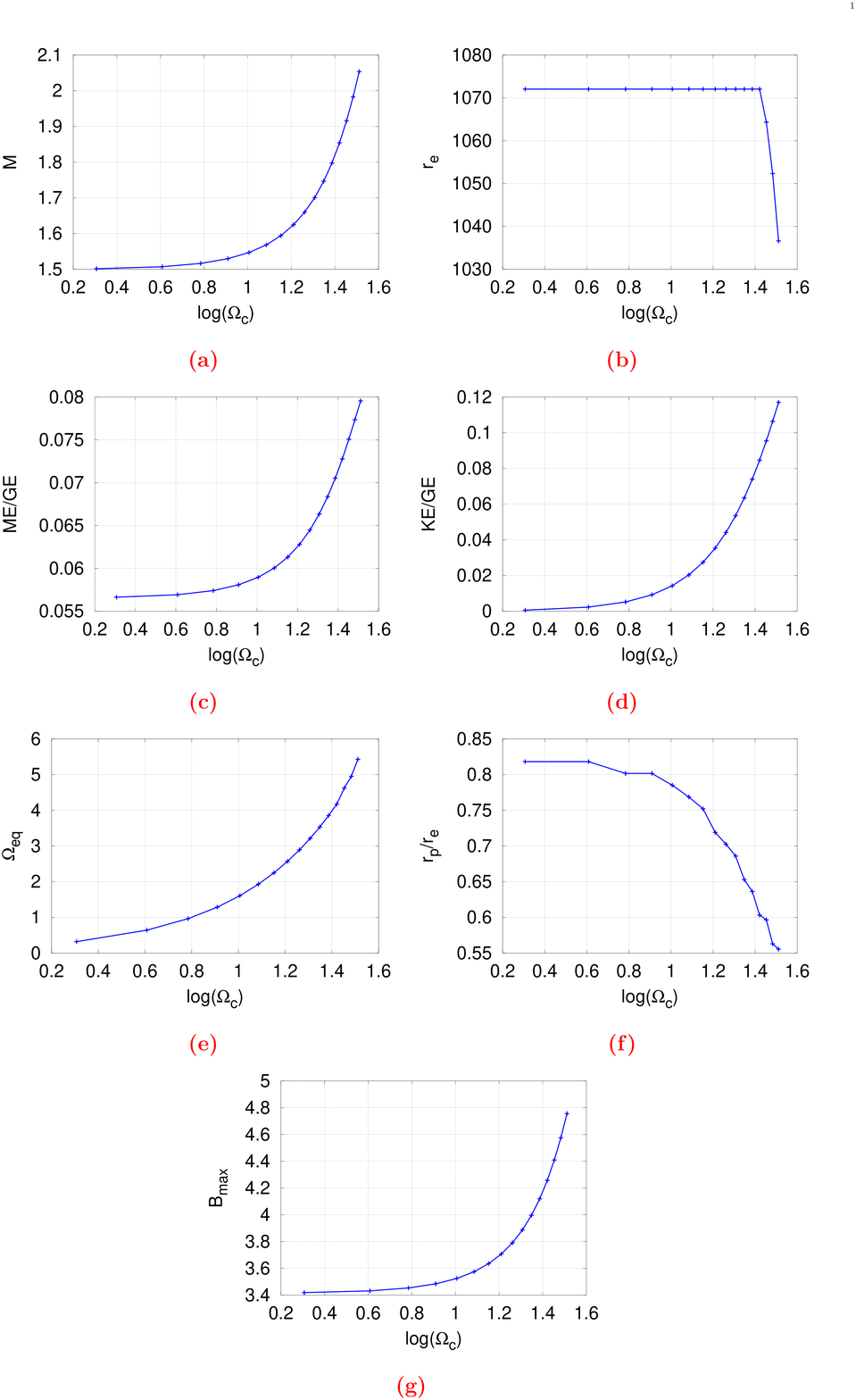}
\caption[Graphs: Sequence 9]{Differentially rotating sequence with changing $\Omega_c$ and
fixed initial poloidal $B_{max}=3.1$. The panels depict 
(a) mass, (b) equatorial radius, (c) ME/GE, (d) KE/GE, (e) equatorial angular velocity, 
(f) $r_p/r_e$, (g) $B_{max}$,  
as functions of the central rotation rate. 
}
\label{polO}
\end{figure}

We note first that for self-consistency, $B_{max}$, fixed initially, again increases from a value of $3.4$,
 in the non-rotating configuration, to $4.8$, for an equatorial rotation 
period of about 1.15 seconds, as shown in Fig. \ref{polO}. The radius changes little, due 
to a cancellation of the effects of rotation and poloidal magnetic field -- the tendency 
of rotation to increase $r_e$, and that of the poloidal magnetic field to decrease 
the radius, almost exactly balance each other in the range of parameters that we 
study. 
%Surprisingly, for larger rotation rates, the radius finally shows a decreasing 
%trend; this is however probably due to the associated increase in the field strength,
%which we allow to be as much as required for the self-consistent solution.
The mass increases by $\sim 37\%$, from $1.502$ to $2.054$, as shown by
Table 9. The mean 
density increases by a factor of 2.3 from the non-rotating, non-magnetised case, reflecting the drastic increase in central condensation, and decrease in size. The formation of polar hollows is apparently hindered by the poloidal 
magnetic field, shown in Fig. \ref{fig:polO}.

Figure \ref{polO} furthermore shows that the equatorial rotation rate increases as $\Omega_c$ increases --- larger rates being allowed due to the smaller radius. ME/GE increases from 0.06 to 0.08 
along the sequence. However, at the higher values of $\Omega_c$, KE/GE overtakes ME/GE, suggesting rotationally dominated effects. This is in contrast to the observed decreasing trend in radius, and the increasing central condensation. We speculate that this may be a purely GR effect, analogous to the case of a Kerr black hole, where increased spin results in more compact event horizons. Since KE/GE is greater than ME/GE at high masses, the mass increase must be more due to rotation than magnetic field effects. Nevertheless, Fig. \ref{polO}(f) reveals some unphysical resolution problem
(with discrete behaviour).

% float containing table and contour plots
% corresponding XNS omg_c - 1e-5, 6e-5, 12e-5, 16e-5
\begin{table}
\centering
\resizebox{\linewidth}{!}{
\begin{tabular}{|c|c|c|c|c|c|c|c|}
\hline 
$\Omega_c$  & $M$ & $r_e$ & $\Omega_{eq}$ & $B_{max}$ & KE/GE & ME/GE & $r_p/r_e$\tabularnewline
\hline 
\hline 
2.028 & 1.502 & 1072 & 0.322 & 3.419 & $6\times 10^{-4}$ & 0.057 & 0.818 \tabularnewline
\hline 
12.168 & 1.568 & 1072 & 1.929 & 3.574 & 0.020 & 0.06 & 0.769 \tabularnewline
\hline 
24.336 & 1.798 & 1072 & 3.854 & 4.119 & 0.074 & 0.071 & 0.636 \tabularnewline
\hline 
32.448 & 2.054 & 1037 & 5.429 & 4.755 & 0.117 & 0.080 & 0.556 \tabularnewline
\hline 
\end{tabular}}
\caption{Differentially rotating configurations with purely poloidal magnetic field with 
changing $\Omega_c$ and initial $B_{max}=3.1$ fixed.}
\end{table}

% stars - contour plots of log-density

% graphs

\subsubsection{Sequence with fixed $\Omega_c$ constructed by varying poloidal $B_{max}$}
\begin{figure}
\centering
\includegraphics[width=\linewidth]{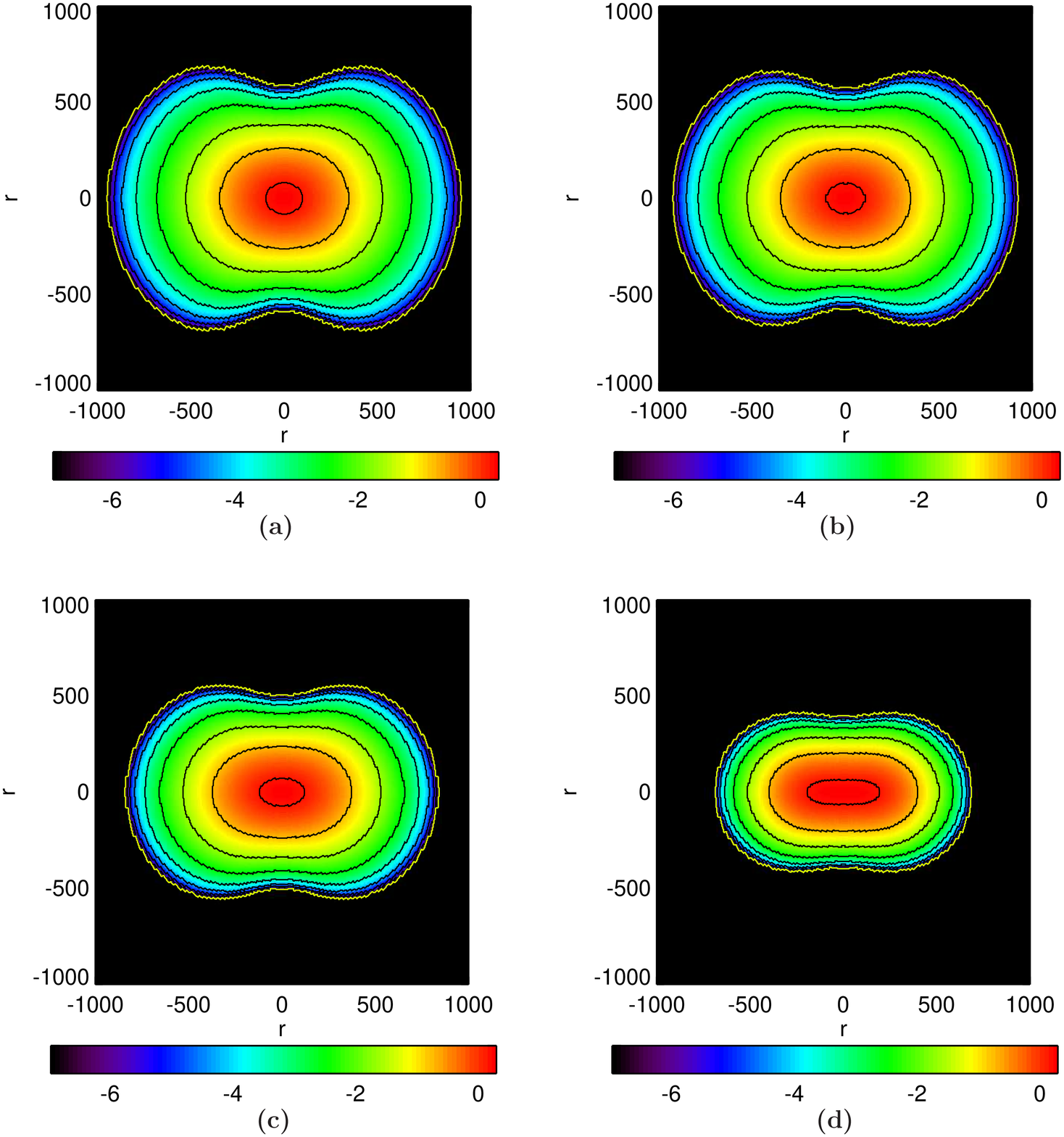}
\caption[Sequence 10: Differentially rotating configurations with purely poloidal magnetic field, with changing $B_{max}$ but fixed field profile]{Sequence of differentially 
rotating configurations with a purely poloidal magnetic field with changing $B_{max}$ 
and $\Omega_c=30.42$ fixed. The panels are contour plots of 
log$\left(\frac{\rho}{\rho_0}\right)$ corresponding to the $B_{max}$ values (a) 0, 
(b) $0.992$, (c) $2.717$, (d) $4.967$. 
The corresponding physical quantities are listed in Table 10.  }
\label{fig:polB}
\end{figure}

\begin{figure}
\centering
\includegraphics[width=\linewidth]{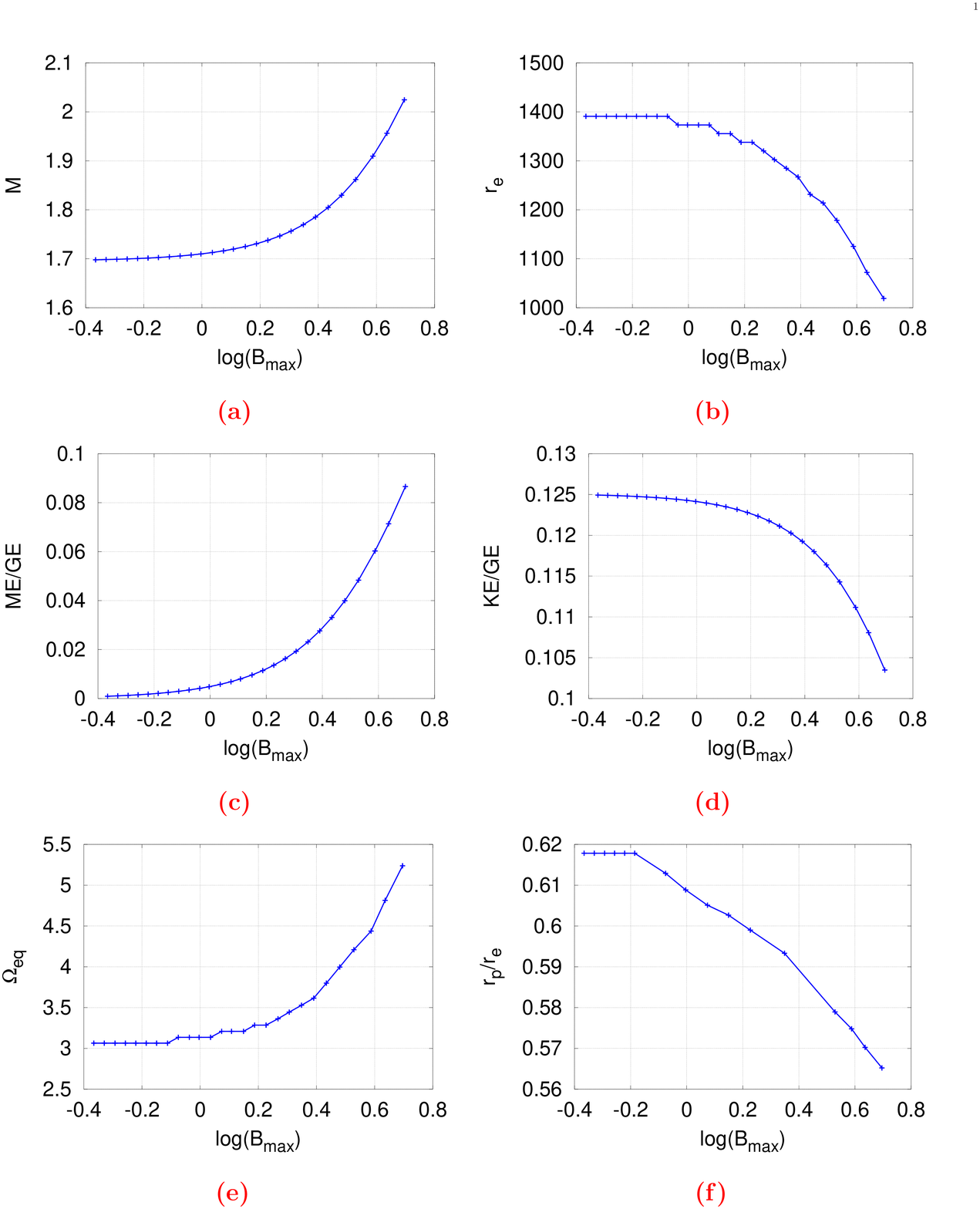}
\caption[Graphs: Sequence 10]{Differentially rotating sequence with changing poloidal 
$B_{max}$ and $\Omega_c=30.42$ fixed. The panels depict 
(a) mass, (b) equatorial radius, (c) ME/GE, (d) KE/GE, (e) equatorial angular velocity, 
(f) $r_p/r_e$,
as functions of the maximum magnetic field. }
\label{polB}
\end{figure}

Figure \ref{fig:polB} and Table 10 show how the increase in poloidal $B_{max}$ results in a decrease of both $r_e$ and $r_p$, counteracting the effect of rotation. The polar hollows clearly become smaller, and we speculate that they may be entirely absent for high enough field strengths (which we were unable to compute due to convergence problems), if other instabilities do not disrupt the configuration. Figure \ref{polB} (with some issues of resolution 
in Figs. \ref{polB} (b) and (e), as mentioned earlier) shows that KE/GE decreases from 0.125 to 
0.103 with the increase of $B_{max}$, indicating the restriction on rotation due to the virial theorem 
(energy conservation) and, hence,
isorotation law. The mass increases by $\sim19.5\%$, from $1.695$ to $2.025$,
 for a field of $B_{max}=4.967$, much less than, however, the increase for 
a toroidal field of similar magnitude. The radius ratio decreases to 0.565, reflecting 
the extremely oblate configurations in Fig. \ref{fig:polB}. The mean density rises by a factor of 2.4, showing the increased degree of central condensation. 
% float containing table and contour plots
% configs given correspond to XNS kbpol = 0, 2.5e-3, 6.24e-3, 9.44e-3
\begin{table}
\centering
\resizebox{\linewidth}{!}{
\begin{tabular}{|c|c|c|c|c|c|c|}
\hline 
$B_{max}$ & $M$ & $r_e$ & $\Omega_{eq}$ & KE/GE & ME/GE & $r_p/r_e$\tabularnewline
\hline 
\hline 
0 & 1.695 & 1409 & 2.996 & 0.125 & 0 & 0.610 \tabularnewline
\hline 
0.992 & 1.710 & 1373 & 3.135 & 0.124 & 0.005 & 0.613 \tabularnewline
\hline 
2.717 & 1.805 & 1231 & 3.800 & 0.118 & 0.033 & 0.597 \tabularnewline
\hline 
4.967 & 2.025 & 1019 & 5.238 & 0.103 & 0.087 & 0.565 \tabularnewline
\hline 
\end{tabular}}
\caption{Differentially rotating configurations with purely poloidal magnetic field, 
with changing $B_{max}$ and $\Omega_c=30.42$ fixed.}
\end{table}

% stars - contour plots of log-density

% graphs

%%% ----------------------------------------------------------------------

\clearpage 	% to enforce placement of all floats

%%%%%%%%%%%%%%%%%%%%%%%%%%%%%%%%%%%%%%%%%%%%%%%%%%%%%%%%%%%%%%%%%%%%%%%%%%%%%%

\section{Exploration of white dwarfs with small radii}

As discussed above, white dwarfs having poloidal magnetic fields can have $r_e$ and $r_p$
much smaller than those having toroidal magnetic fields. The question is, how small could
a white dwarf be in size? In order to explore the answer to such a question,
we consider the following two cases.\\

\begin{figure}
\centering
\includegraphics[width=\linewidth]{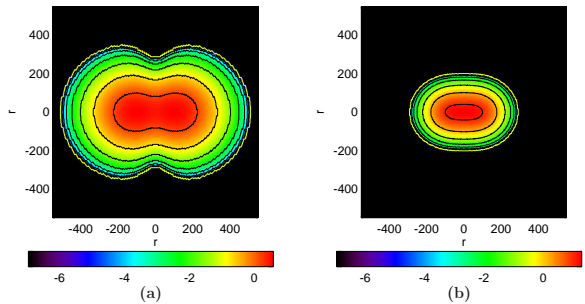}
\caption{Illustration of smaller white dwarfs: The panels are contour plots of 
$\log\left(\frac{\rho}{\rho_0}\right)$ for differentially rotating white dwarfs
with poloidal magnetic fields having (a) $M=1.95$, $r_e=746$, $r_p/r_e=0.564$, and 
(b) $M=1.72$, $r_e=428$, $r_p/r_e=0.689$. For other 
physical quantities, see the description in the text.  }
\label{smallr}
\end{figure}

\noindent (a) A white dwarf with $\rho_c=3\times 10^{10}$ g/cc, $\Omega_c=111.5$, $\Omega_{eq}=4.16$, $A^2=10^4$,
$B_{max}=7.22$, giving rise to $M=1.95$, $r_e=746$, $r_p/r_e=0.564$ when KE/GE=0.1033, ME/GE=0.071;
shown in Fig. \ref{smallr}a.\\ \\
\noindent (b) A white dwarf with $\rho_c=2\times 10^{11}$ g/cc, $\Omega_c=40.56$, $\Omega_{eq}=4.25$, $A^2=10^4$,
$B_{max}=25.2$, giving rise to $M=1.72$, $r_e=428$, $r_p/r_e=0.689$ when KE/GE=0.011, ME/GE=0.1133;
shown in Fig. \ref{smallr}b.\\

Figures \ref{smallr}a and b, depicting the density isocontours of the above mentioned white dwarfs, 
indeed demonstrate the possible existence of white dwarfs with much smaller radii compared to their
conventional counterparts. They could be, in fact, potential candidates for soft gamma-ray repeaters 
and anomalous X-ray pulsars with much smaller surface magnetic fields than that needed for a 
neutron star based model (magnetar). Now, one has to check the stability of such white dwarfs,
in particular the one in Fig. \ref{smallr}b having higher $\rho_c$, even though both models have acceptably small KE/GE and ME/GE. However, note that the threshold densities for the onset of
several kinds of instability in general relativity in the presence of a high magnetic field would
be expected to be different (probably higher) than those in Newtonian mechanics with a low magnetic field. 
Nevertheless, the value of $\Gamma$ for the white dwarf in Fig. \ref{smallr}b need not be 
$4/3$ throughout due to its higher $B_{max}$. In spite of these cautionary notes, the above 
exploration argues for the existence of white dwarfs having $r_e$ only an order or order and half
magnitude higher than that of typical neutron star. Hence, they can be very good storage of a large amount of
magnetic energy.

\section{Summary and Conclusions}
Magnetic field and rotation are important elements in the theory and phenomenology of white dwarfs, and realistic models must endeavour to include them. Differential rotation is expected to be a common feature in all kinds of star. As we have seen, it can significantly modify the structure of the star, and can cause the formation of concavities or hollow regions near the poles, in special cases. The geometry of the magnetic field also plays a crucial role in determining the structure of the white dwarf, and toroidal and poloidal geometries introduce two entirely different kinds of deformation. We have computed sequences of equilibrium configurations exhibiting magnetic fields of these two geometries, along with the inclusion of both uniform and differential rotations. We have adopted a GRMHD framework that has extensively been used by many authors to investigate neutron star structure, implemented numerically in the {\it XNS} code. We have modified {\it XNS} to work for white dwarfs, and the use of this code allows a better treatment of the non-linear Einstein equations in the presence of strong magnetic fields than in previous perturbative or Newtonian calculations (\citealt{pili}).

%%%%%%%%%%%%%%%%%%%%%%%%%%%%%%%%%%%%%%%%
%(\emph{are there any for WDs? KEH?} ).%
%%%%%%%%%%%%%%%%%%%%%%%%%%%%%%%%%%%%%%%%

\vspace{0.3cm}
For a white dwarf of fixed central density $\approx 2\times 10^{10}$ g/cc, and magnetic fields of magnitude $\lesssim 5\times 10^{14}$G, we summarise below the key results from our parameter space study.
\vspace{-0.2cm}

\begin{itemize}
\item Differential rotation in the absence of magnetic fields can result in stars with masses of $1.4 - 1.8\, M_\odot$, having equatorial radii in the range $1200 - 1450$km. The mass, equatorial 
radius, and the ratio of kinetic to gravitational energies increase 
with the increase of central (and, therefore, surface) angular velocity, apparently without an intrinsic bound. The largest surface equatorial angular velocities considered correspond to rotation periods of $1-2$ seconds. There is a marginal decrease in the degree of central condensation, with mean density decreasing by about 30\%. These stars show an oblate deformation, with the ratio of polar to equatorial radius dipping as low as 0.6. In special cases, when the central angular velocity is greater than that of the corresponding uniformly rotating configuration at the mass-shedding limit, concavities or hollow regions develop around the poles. Slow variation of angular velocity in the radial direction, as compared to steep variation, causes larger deformations (always oblate), but less increase in mass. However, such flat rotation profiles cause smaller polar hollows, with the hollows disappearing altogether when the rotation profile approaches rigid body (uniform) rotation.  
\item Toroidal magnetic fields in the absence of rotation give rise to stars in the mass range $1.4-2.3\,M_\odot$, 
and equatorial radius range $1200-2330$ km. Mass, radius and magnetic to gravitational energies ratio increase with the increase of maximum value of the magnetic field, which we have allowed to be as large as $3.4\times 10^{14}$G. The surface of these stars deviates very little from a spherical shape, as indicated by low values, $<1.08$, of the ratio of polar to equatorial radii. However, the matter distribution in the interior shows extreme prolate deformation. The mean density falls by almost a factor of 10, indicating bloated outer regions of low density. When differential rotation is also present, the mass range is enhanced to $1.4-3.1\,M_\odot$, with radii increasing to nearly 3300 km. The corresponding 
magnetic field has maximum magnitude of $3.6\times 10^{14}$G and the corresponding surface equatorial rotation 
period is about $10$ seconds. Mass and radius both increase when the maximum field or the central 
angular velocity is increased, apparently without bound. However, peaks can be seen in the ratio of kinetic 
to gravitational
energy and the surface equatorial angular velocity as functions of maximum field and central rotation rate respectively, indicating the possible onset of some kind of instability. The interior regions continue to show a prolate deformation, while the exterior tends to be oblate. When the differential rotation also causes polar hollows, the ratio of polar to equatorial radii falls to about 0.6. Differential rotation induced polar hollows are aggravated by toroidal magnetic fields.
\item Poloidal magnetic fields in the absence of rotation give rise to stars in the mass range $1.4-1.6\,M_\odot$, 
and equatorial radius range $1200-1000$ km. Mass and the ratio of magnetic to gravitational energies increase 
with the increase of maximum value of the magnetic field, while the radius decreases. We have generally 
taken fields with maximum magnitude as high as $3.9\times 10^{14}$G. Both the surface and the matter distribution 
show oblate deformation, with values of the ratio of polar to equatorial radii in the range $1-0.79$. 
The mean density increases by nearly 80\%, indicating highly centrally condensed and compact structures. 
When differential rotation is also present, the mass range is enhanced to $1.4-2.1\,M_\odot$, and the equatorial 
radius range to $1400-1000$ km. The corresponding magnetic field has maximum magnitude of $5\times 10^{14}$G, 
and the corresponding surface equatorial rotation period is about $2$ seconds. There is an extreme oblate 
deformation, with the ratio of polar to equatorial radius decreasing to about 0.55. The maximum value of the 
magnetic field generally increases with the increase in rotation rate, when the parameters determining the field are 
held fixed. Mass increases with both the maximum field and the angular velocity, apparently without bound. 
However, the radius can either decrease or increase, depending on whether the field or the rotation effects 
dominate. Although both poloidal fields and rotation act in the same way with regard to deformation, namely 
by causing oblateness, the rotation causes increased size while poloidal field causes decreased size. If polar 
hollows are present, increasing the magnitude of the field results in a smoothening out of the polar hollows, 
due to the possible cancellation of centrifugal pressure and magnetic pressure. Furthermore, by increasing the
central density and maximum magnetic field, super-Chandrasekhar white dwarfs could be shown to be much smaller with 
the equatorial radius $400-750$ km.  
\end{itemize}

The points noted above lead us to conclude that white dwarfs with highly super-Chandrasekhar masses, i.e. $M>M_{Ch}=1.44\,M_\odot$, are possible when rotation and magnetic fields are taken into consideration.

Anticipating the existence of an upper mass limit for magnetised and rotating white dwarfs (\citealt{prl13}), we propose that the equilibrium solutions we have constructed here could be candidates for the progenitors of over-luminous,
peculiar type Ia supernovae. The explanation of such supernovae requires highly super-Chandrasekhar white dwarfs with masses in the range of $2.1-2.8\,M_\odot$, and many of the configurations we have constructed in this work are within this mass range.

%The deformations due to rotation and magnetic fields also have important implications. The distinction between an apparent oblate shape (surface), but prolate matter distribution, can crucially alter the emission of gravitational waves by rotating compact objects, and can have observational consequences once gravitational wave detectors begin to record data at higher sensitivities. 

%There are seveal open problems in the area of magnetised and rotating compact objects in GR. We have raised many questions of our own in the course of this work, and plan to take up the questions regarding the stability criteria for such objects, the possibility of fission of differentially rotating structures showing polar hollows, and the investigation of more differential rotation profiles, in a future work. 

%%%%%%%%%%%%%%%%%%%%%%%%%%%%%%%%%%%%%%%%%%%%%%%%%%%%%%%%%%%%%%%%%%%%%%%%%%%%%%

\section*{Acknowledgments}
We thank J. P. Ostriker for continuous encouragement, discussion and helping us to understand 
his work in this field and compare that with our work. 
We are also truly grateful to 
N. Bucciantini and A.G. Pili for their extremely helpful  inputs 
regarding the execution of the {\it XNS} code. Finally, we thank Upasana Das for discussions and help with execution of the {\it XNS} code. S.S. thanks KVPY, India for financial support.
%B.M. acknowledges partial support through research Grant No. ISRO/RES/2/367/10-11. 

%%%%%%%%%%%%%%%%%%%%%%%%%%%%%%%%%%%%%%%%%%%%%%%%%%%%%%%%%%%%%%%%%%%%%%%%%%%%%%

\bibliographystyle{apa}

\label{lastpage}

\end{document}